\documentclass[12pt,a4paper]{article}
\usepackage[textheight=23cm,textwidth=15cm]{geometry}
\usepackage{amsmath}
\usepackage{graphicx}
\usepackage[american]{babel}
\usepackage{here}
\usepackage[maxauthors=100, articletitle=true, super=true]{achemso}
\usepackage{multirow}

\renewcommand\arraystretch{1.5}

\usepackage{url} % necessary?
\usepackage{xcolor} % only temporary for the comments

\begin{document}

\renewcommand\thesection{\Roman{section}}
\renewcommand\thesubsection{\Roman{section}.\arabic{subsection}}
\renewcommand{\labelitemii}{$\longrightarrow$}

\begin{center}

{\LARGE\bf
  Self-Parametrizing System-Focused\\[0.3cm]Atomistic Models
}

\vspace{1cm}

\renewcommand*{\thefootnote}{\fnsymbol{footnote}}

{\large
Christoph Brunken
and
Markus Reiher\footnote{Corresponding author; e-mail: markus.reiher@phys.chem.ethz.ch}
}\\[4ex]

\renewcommand*{\thefootnote}{\arabic{footnote}}
\setcounter{footnote}{0}

ETH Zurich, Laboratory for Physical Chemistry, Vladimir-Prelog-Weg 2,\\ 
8093 Zurich, Switzerland \\

\vspace{.5cm}

December 13, 2019

\vspace{.43cm}

\textbf{Abstract}
\end{center}
\vspace*{-.41cm}
{\small

Computational studies of chemical reactions in complex environments such as proteins, nanostructures, or on surfaces
require accurate and efficient atomistic models applicable to the nanometer scale.
In general, an accurate parametrization of the atomistic entities will not be available for arbitrary system classes, but demands a fast automated system-focused parametrization procedure
to be quickly applicable, reliable, flexible, and reproducible.
Here, we develop and combine an automatically parametrizable quantum chemically derived molecular mechanics model with machine-learned corrections
under autonomous uncertainty quantification and refinement.
Our approach first generates an accurate, physically motivated model from a minimum energy structure and its corresponding Hessian matrix
by a partial Hessian fitting procedure of the force constants.
This model is then the starting point to generate a large number of configurations for which additional off-minimum reference data can be evaluated on the fly.
A $\Delta$-machine learning model is trained on these data to provide a correction to energies and forces including uncertainty estimates.
During the procedure, the flexibility of the machine learning model is tailored to the amount of available training data.
The parametrization of large systems is enabled by a fragmentation approach.
Due to their modular nature, all model construction steps allow for model improvement in a rolling fashion.
Our approach may also be employed for the generation of system-focused electrostatic molecular mechanics embedding environments in a
quantum-mechanical/molecular-mechanical hybrid model for arbitrary atomistic structures at the nanoscale.
}

\newpage
\section{Introduction}
\label{sec:introduction}

Complex molecular processes and reaction chemistry can be found in huge molecular frameworks
ranging, for instance, from protein machinery in cells to large heterogeneous structures on surfaces and to metal-organic frameworks~\cite{farrusseng11}.
Their sheer size combined with a lack of symmetries that could be exploited renders a straightforward quantum chemical approach unfeasible and a mixed model will be required.
However, in such a multiscale modeling approach, a parametrized non-quantum part is in general not readily available for arbitrary system classes.
Therefore, a challenge in the field is to develop sophisticated and at the same time efficient methods
for the modeling of arbitrary atomistic structures at the nanoscale.

An additional technical challenge is the generation of sensible starting structures for nanoscale systems.
For proteins, structures are typically sampled from molecular dynamics (MD) simulations starting from crystallographic data provided by the Protein Data Bank~(PDB)~\cite{bernstein77},
solvating water molecules are added according to heuristic rules~\cite{humphrey96}
and the protonation states of residues are assigned based on pH values~\cite{word99}.
However, there exist no fully integrated and standardized workflows for these procedures in general
and they become unfeasible for nanoscale structures for which no such structural database exists.

If then chemical reactions in such nanostructures shall be described, one usually resorts
to hybrid quantum mechanical/molecular-mechanical (QM/MM) methods,
in which the MM environment is modeled by a classical force field and the QM region allows for the description of the bond breaking processes~\cite{senn06, groenhof13}.
Although the QM/MM approach\cite{warshel76} has already been introduced in 1976 and its importance was recognized by a Nobel prize in 2013~\cite{thiel13},
a variety of challenges still hampers its truly widespread application.
Setting up QM/MM models for complex systems remains an inefficient and tedious task.
The separation of the system into QM and MM regions requires significant manual interference in the model construction process.
Moreover, it was demonstrated that the convergence of calculated properties can be very slow with respect to the size of the QM region~\cite{kulik16, rossbach17}.
Furthermore, it is desirable to have a flexible and adaptive definition of quantum mechanically treated atoms~\cite{heyden07, bulo09, mones15, zheng16, duster17}.

Here, we consider the first challenge, namely to provide efficient molecular models for large atomistic structures.
We start from quantum mechanical data in order to later straightforwardly extent to QM/MM hybrid models.
In a natural fashion, our system-focused approach can adapt to incompatibilities between the QM and MM representations in these methods (e.g., to avoid overpolarization effects), which was identified as a critical issue in
QM/MM calculations.~\cite{muller95, heimdal12, fox13, cave14, mikulskis14, hudson15, genheden15, sampson15, ryde17, kearns17, olsson17, boresch17, caldararu18, hudson18, konig18_a, konig18_b, wang19}
This work therefore focuses on the bottom-up construction of chemical force fields from parameter-free quantum chemical calculations
on small molecules or molecular fragments.

There exists a large variety of force fields with different properties and specific areas of application\cite{mackerell04}.
For the modeling of peptides and proteins alone one may choose from over twenty different classes of force fields\cite{ponder03}.
The most established ones are GROMOS~\cite{scott99, schuler01, oostenbrink04, schmid11}, AMBER~\cite{cornell95, wang04},
CHARMM~\cite{brooks83, mackerell98, vanommeslaeghe10, pastor11, huang13},
OPLS~\cite{jorgensen88} and MMFF~\cite{halgren96}. These claim to be universally transferable within the substance
classes for which they have been parametrized. However, they do not represent generally applicable parametrizations for
a wide range of chemical space. To what extent the transferability of the parameters can
be safely assumed is somewhat unclear, because system-focused uncertainty quantification is hardly available.
Furthermore, introducing non-standard structures (e.g., those containing $d$-metal atoms) into the system, usually
creates massive practical complications, because non-standard parameters have to be
generated and uncommon bonding patterns may not be supported.

Obviously, general applicability of an atomistic model usually points toward approximate electronic structure methods.
The latter are agnostic
with respect to atom types in the molecule and instead invoke approximations for the interaction of the charged elementary entities
(i.e., electrons and atomic nuclei), most importantly for the electron-electron interaction. Because of these approximations even such universal
models are not truly generally applicable, and even not necessarily uniquely defined (see, for instance, the myriad of equally reliable 
density functionals that can be constructed from some given reference data~\cite{mardirossian15, mardirossian17})
and their transferability to unseen systems is, in general, neither guaranteed nor known, although attempts were made to change this
situation~\cite{mortensen05, wellendorff12, wellendorff14, pernot15, simm16, simm18_gp, pernot18}.

The derivation of fast, but transferable atomistic electronic structure models applicable
to molecules composed of any elements from the periodic table has been driven by Grimme and co-workers to an extreme degree~\cite{grimme17}. They
significantly extended and parametrized~\cite{pracht19} non-self-consistent density-functional tight-binding methods~\cite{seifert96} 
to obtain universal and efficient atomistic models. However, the transferability of these boxed models is difficult to assess. Instead,
uncertainty quantification~\cite{simm17error} and error reduction for the specific cases under consideration would be desirable, as already exemplified
for general physical models~\cite{proppe17} and dispersion interactions~\cite{weymuth18, proppe19}.
Compared to such non-iterative one-shot electronic structure models, MM approaches still
have the advantage of being orders of magnitude faster than the fastest non-self-consistent tight-binding approaches and do not
suffer from the formally cubic scaling of the diagonalization of large matrices.

It is, therefore, desirable to develop a black-box modeling approach with simple and therefore fast to evaluate 
model potentials for interacting subentities with built-in refinement and no restrictions on the type of molecular system. 
This approach must therefore be based on first-principles methods of quantum chemistry, which can be applied to molecular systems of any
composition of elements from the periodic table. 
A system focus requires that a molecular model is parametrized automatically in a theoretically consistent manner (i) with
system-specific uncertainty measures that eventually lead to a rolling refinement of the model
and (ii) within a parametrization time that is negligible compared to the time scales required for subsequent simulations with this model
in order for it to be practical.

While an approach that derives an atomistic MM model for a specific nanoscale molecular assembly would be system-focused by construction,
we emphasize that this system specificity of the model will eventually become mandatory if quantitative results shall be obtained
with the model. For instance, in situations where energy differences enter exponential expressions, as in chemical rate constants,
qualitative agreement of the atomistic model with experimental data will no longer be sufficient for predictive work. 
However, as soon as many such system-focused models become available, one may apply advanced machine learning schemes to identify transferable
components and parameters that eventually enhance future model construction.

Approaches to derive force fields from quantum chemical data have already been proposed in the literature.
Recent examples are Grimme's Quantum-Mechanically Derived Force Field (QMDFF)~\cite{grimme14} with its empirical valence bond extension EVB-QMDFF~\cite{hartke15} 
and the QuickFF~\cite{vanduyfhuys15} automated parametrization scheme by Van Speybroeck and coworkers.
In both models, the force field parameters are generated from quantum mechanically calculated reference data to obtain a system-specific model for a given molecular system.
The parametrizations require reference information on the molecular structure of the system, second derivatives of its energy with respect to nuclear positions, the Hessian matrix,
as well as atomic partial charges and covalent bond orders. These data can be produced with any electronic structure model.
The essential advantages of these approaches are (i) that they are constructed to accurately reproduce the potential energy surface of its reference QM
method close to the minimum energy structure and (ii) that they can, in principle, treat any reasonable covalent structure.
However, a force field parametrization based on the electronic energy landscape close to the equilibrium structure
is not expected to be able to cope with molecular configurations representing more distant regions on the potential energy surface~(PES).
Also, the scaling of these approaches to large nanoscale systems might become an issue.
Consequently, algorithms to automatically build a model for large biomolecules from force field parameters of molecular fragments by graph matching have been proposed.~\cite{allison19}

To address the latter issue, we set out from our work on uncertainty quantification (see references given above) and
propose a new strategy to obtain self-parametrizing system-focused atomistic models (SFAMs), designed to be equipped with uncertainty quantification.
For semi-empirical quantum chemical methods, a similar intent led us to the development of the correction inheritance to semi-empirics (CISE) approach~\cite{husch18},
a system-focused model, which allows to apply a correction matrix to semi-empirical calculations obtained from a reference method.
This matrix is transferable for moderate structure modifications and can be employed until severe structural changes require new reference calculations
on the fly, which can be measured, for instance, by kriging~\cite{simm18_gp}.

\begin{figure}[H]
\begin{center}
\includegraphics[width=0.63\textwidth,trim={0cm 0cm 0cm 0cm},clip]{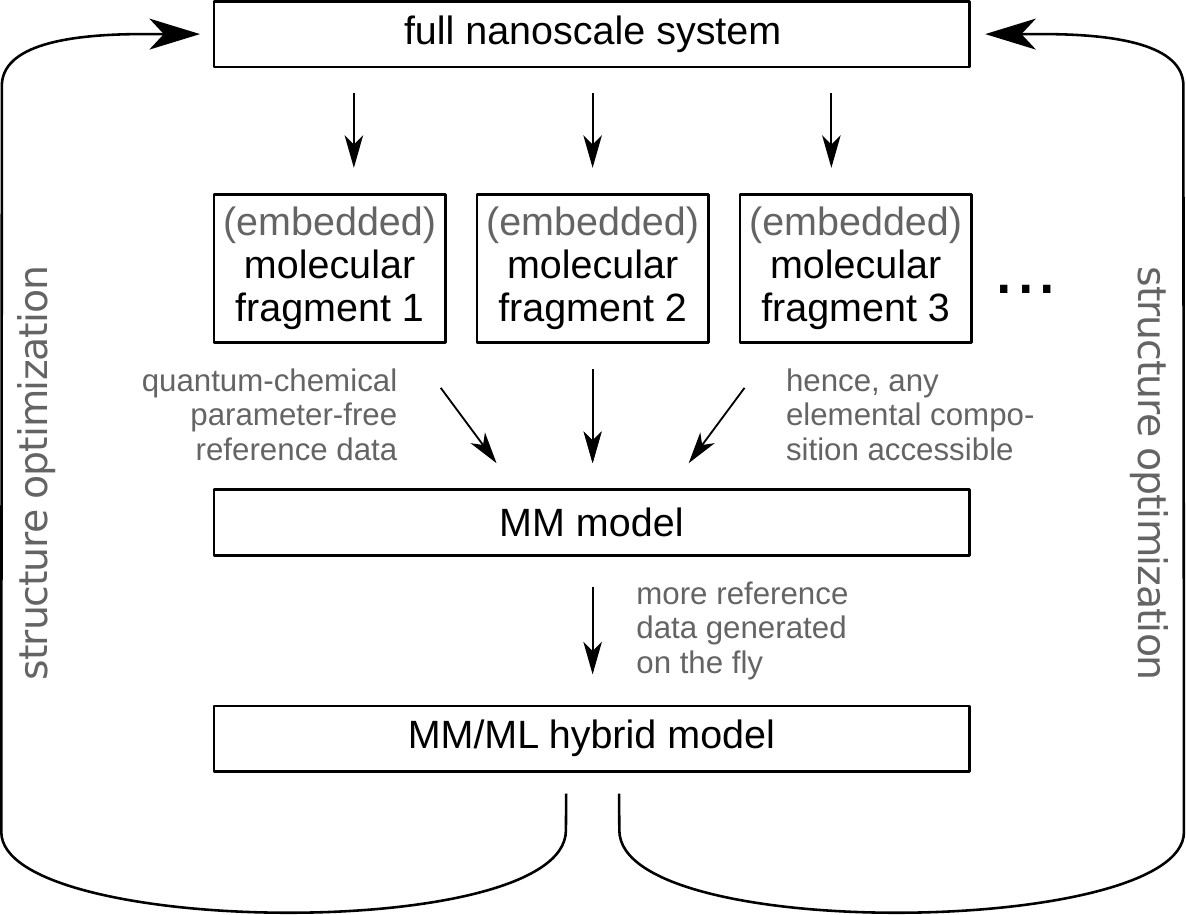}
%trim={<left> <lower> <right> <upper>
\end{center}
\vspace*{-0.25cm}
\caption{\label{fig:proposed_workflow}\small SFAM workflow for generating self-parametrizing system-focused models.
An MM model is obtained from an automated fitting procedure to first-principles reference data generated for automatically dissected
large molecular structures broken down to reasonably sized fragments (i.e., those for which reference data can be obtained in a
time frame small compared to the overall simulation protocol),
which can then be employed to sample molecular configurations. 
Additional reference data allows to train an ML model on the fly
for gradually increasing accuracy of molecular property calculations including uncertainty quantification.
The model obtained in a rolling fashion can also be applied iteratively to obtain an optimized structure for the full nanoscale system.}
\end{figure}

Due to efforts which combine physically motivated models with data-driven approaches~\cite{botu16, li17, ramprasad17, kulakova17, bartok17, deringer17, bartok18, reichstein19, amabilino19},
machine learning (ML) has been established as a universal tool to estimate corrections for a base model with respect to a more accurate reference model.
This was introduced to computational chemistry by von Lilienfeld and coworkers and is known as $\Delta$-machine learning.~\cite{ramakrishnan15}
For our approach, we combine the automated parametrization of a classical force field with $\Delta$-ML as illustrated in Fig.~\ref{fig:proposed_workflow}.
An atomistic MM model is automatically parametrized for any given molecular structure based on quantum chemical data. This allows for a fast generation of a base model
for arbitrary molecular structures composed of any element of the periodic table due to the first-principles basis,
for which we demand a sufficiently accurate description of the PES close to the equilibrium structure
as well as a qualitatively correct description of the main features of the overall PES.
However, we emphasize that any quantum chemical method may be applied ranging from semi-empirical approaches~\cite{husch18_review}
to accurate explicitly correlated coupled cluster methods with sufficiently high excitation rank~\cite{ma18} or to 
advanced multi-configuration approaches~\cite{baiardi19}.

This strategy enables us to perform calculations with early versions of our model, i.e., during the model construction process where,
owing to a lack of data, the model will be affected by large uncertainties.
This will require less initial reference data than demanded by common ML-only approaches for molecular force fields~\cite{chmiela17, chmiela18}. 
As more molecular configurations are sampled, for instance, by MD simulations or by systematic search~\cite{rdkit, riniker15}
as part of an automated reaction network exploration~\cite{simm17, simm18, unsleber19},
reference data can then be acquired on the fly~\cite{csanyi04} for so far unseen molecular configurations.
Subsequently, a $\Delta$-ML model is trained on the incoming data to stepwise increase the overall model accuracy and to reduce uncertainty estimates for the hybrid MM/ML model.
The atomistic model for the entire nanoscale system can be applied to optimize the initial atomistic structure,
which might have been chosen rather arbitrarily or on the basis of uncertainty-affected experimental data.
By this procedure, we are able to obtain such nanoscale structures iteratively.

In the following, we first introduce our automated MM parametrization scheme based on a partial Hessian fit
and outline how it can be assembled from multiple molecular fragments obtained by dissecting a large molecular structure.
Then, a $\Delta$-ML approach to improve on this base model on the fly with uncertainty quantification is described.

%%%%%%%%%%%%%%%%%%%% AUTOMATED MM PARAMETRIZATION %%%%%%%%%%%%%%%%%%%%
\section{Automated parametrization of SFAM}
\label{sec:automated_mm_parametrization}

\subsection{Definition of the MM model}
\label{sec:model_definition}

\begin{figure}[H]
\begin{center}
\includegraphics[width=0.5\textwidth,trim={0cm 2.2cm 1cm 2.2cm},clip]{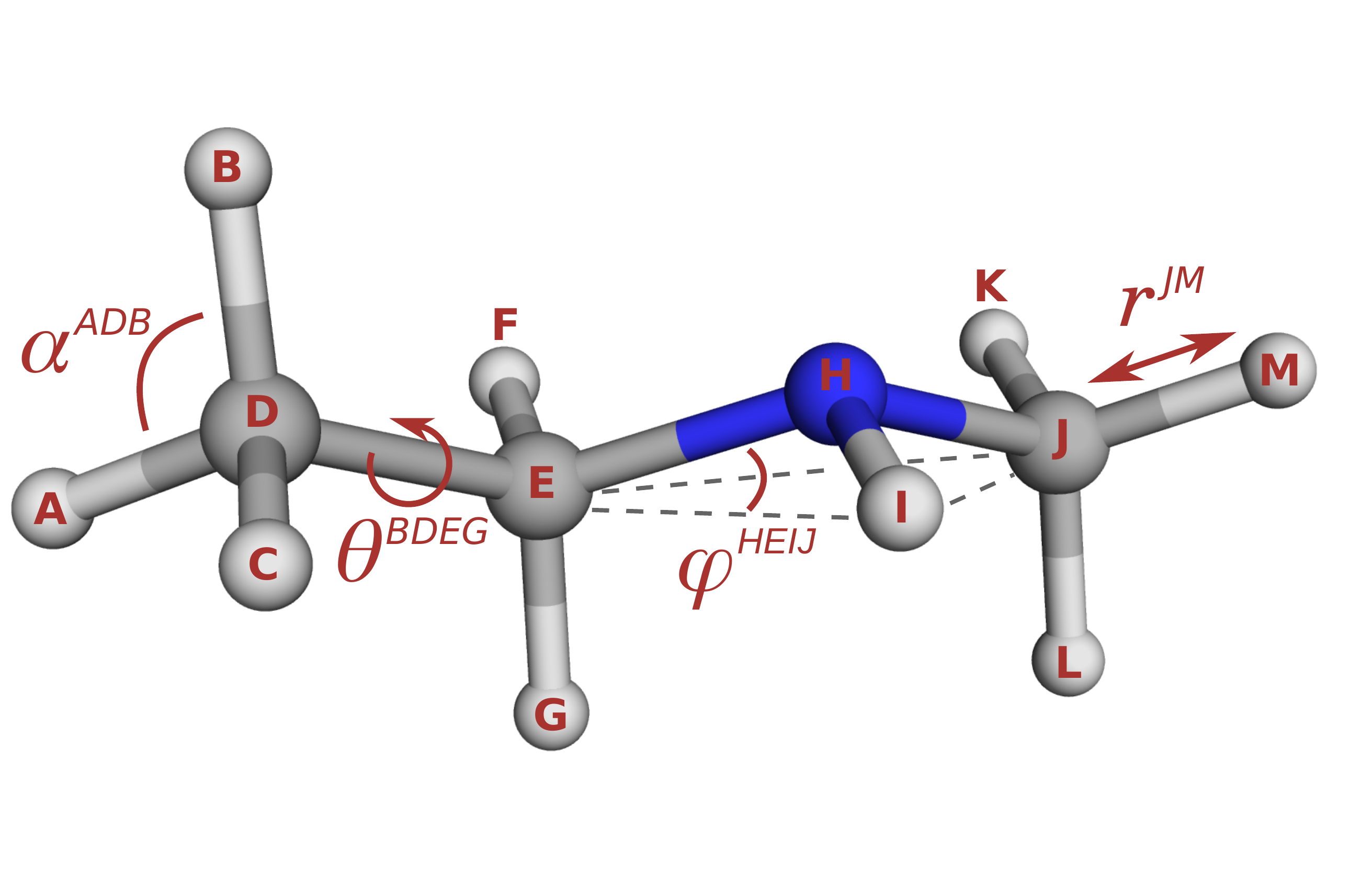}
%trim={<left> <lower> <right> <upper>
\end{center}
\caption{\label{fig:internal_degrees_of_freedom}\small Illustration of the internal degrees of freedom contributing to the covalent
components of the MM potential energy at the example of ethylmethylamine.}
\end{figure}

There exist several approaches in the literature for generating a classical force field based on quantum chemical reference data.~\cite{grimme14, vanduyfhuys15, allen18, hagler15}
As it is common in MM models~\cite{riniker18}, we calculate the total potential energy $E_{\text{MM}}$
as the sum of non-covalent interactions $E_{\text{nb}}$ and covalent contributions $E_{\text{cov}}$,
\begin{equation}
 E_{\text{MM}} = E_{\text{cov}} + E_{\text{nb}} \quad ,
\end{equation}
where the covalent terms are calculated from displacements of the internal degrees of freedom depicted in Fig.~\ref{fig:internal_degrees_of_freedom},
\begin{equation}
 E_{\text{MM}} = E_{r} + E_{\alpha} + E_{\theta} + E_{\varphi} + E_{\text{estat}} + E_{\text{disp}} + E_{\text{rep}} + E_{\text{hb}} \quad , \label{eq:force_field_general}
\end{equation}
including bonds $r$, bond angles $\alpha$, dihedral angles $\theta$ and improper dihedral angles $\varphi$, from their equilibrium values.
Furthermore, Eq.~(\ref{eq:force_field_general}) illustrates that the non-covalent interactions are split into an electrostatic term $E_{\text{estat}}$,
dispersion and Pauli repulsion interactions, $E_{\text{disp}}$ and $E_{\text{rep}}$, respectively, as well as additional corrections for hydrogen bonds $E_{\text{hb}}$.
Following the standard ansatz for a non-reactive force field, we construct our potential energy contributions from bonds and bond angles as harmonic potentials,
\begin{equation}
 E_{r} = \sum_{(A,B)} \frac{1}{2} \, k_r^{AB} \, (r^{AB} - r_0^{AB})^2 \quad , \label{eq:bond_term}
\end{equation}
and
\begin{equation}
 E_{\alpha} = \sum_{(A,B,C)} \frac{1}{2} \, k_\alpha^{ABC} \, (\alpha^{ABC} - \alpha_0^{ABC})^2 \quad , \label{eq:angle_term}
\end{equation}
summed over all bonded atom pairs $(A,B)$ with internuclear distance $r$ and all bonded atom triples $(A,B,C)$ with angle $\alpha$.
The molecular graph's connectivity is, in our case, automatically extracted from quantum chemical population analysis, specifically, 
from Mayer bond orders~\cite{mayer83, mayer86}.
Eq.~(\ref{eq:bond_term})~and~(\ref{eq:angle_term}) contain four parameters, the force constants $k_r$ and $k_\alpha$,
as well as the equilibrium distances $r_0$ and equilibrium angles $\alpha_0$ obtained by quantum chemical structure optimization.
The potential energy of Eq.~(\ref{eq:bond_term}) grows quadratically with increasing displacement and therefore prohibits bond breaking.
However, our system-focused quantum mechanically based parametrization also allows us to build a reactive MM model similar to, e.g., the ReaxFF approach~\cite{vanduin01, senftle16}.

We first focus on a simple MM ansatz in a non-reactive formalism,
because we plan to exploit efficient quantum chemical methods for bond breaking in a QM/MM approach with error-control~\cite{husch18, simm16, simm18_gp} as a second step.
We consider the extension toward reactive force fields a final step in order to counter the computational cost of a large number of QM/MM calculations.

The dihedral-angle contributions are modeled by a single cosine function for each set of four sequentially bonded atoms $(A,B,C,D)$,
\begin{equation}
 E_{\theta} = \sum_{(A,B,C,D)}  V_\theta^{ABCD} \, \left( 1 - \cos \left( n^{ABCD} \theta^{ABCD} - \theta_0^{ABCD} \right) \right) \quad , \label{eq:dihedral_potential}
\end{equation}
where $V_\theta$ is the half barrier height of the dihedral angles, $n$ is the periodicity of the potential energy corresponding to the number of its minima and $\theta_0$ is a phase shift.

We note that some force fields such as AMBER~\cite{cornell95} and QMDFF~\cite{grimme14} implement a Fourier series to describe dihedral-angle
distortions instead,
which leads to a more flexible and, hence, potentially more accurate potential.~\cite{hagler15}
However, we refrain from such extensions as we intend to grasp deficiencies of the simple model by subsequent machine learning as discussed in part~\ref{sec:ML} of this work.
Furthermore, we found that such an approach is less stable toward our parametrization procedure described in section~\ref{sec:mm_parameter_optimization},
because we parametrize solely on local quantum mechanical information at the minimum energy structure, which is often not sufficient to generate a satisfactory fit to a more complex potential.
In QMDFF, this problem is avoided by fitting dihedral potentials to explicitly calculated $2\pi$ rotational potentials with a minimal valence basis set tight-binding Hamiltonian.~\cite{grimme14}
However, more advanced dihedral potentials may be beneficial to accurately describe long-range structural arrangements in nanoscale structures, such as
the secondary structure of biopolymers~\cite{mackerell04_md}.
Such a refinement will be beneficial for SFAM as well and may be integrated into the iterative structure generation process (see Fig.~\ref{fig:proposed_workflow}). It will require additional first-principles
information, which may in many cases simply be taken from the D3 model in conjunction with a fast electronic structure model such as Grimme's GFN-xTB~\cite{grimme17}
applied to a significantly extended fragment structure.

Improper dihedral potentials serve two purposes in molecular mechanics.~\cite{riniker18}
First, planar fragments $(X,A,B,C)$ benefit from such an auxiliary potential in order to more accurately preserve planarity
($X$ denotes the central atom of the improper dihedral angle).
Second, the inversion barrier of trigonal pyramidal fragments needs to be accurately described.
We apply two different potentials for the two separate cases, namely a harmonic potential for equilibrium out-of-plane angles $\varphi_0$ smaller than a threshold value $\tilde{\varphi}$
and a double-well potential for large equilibrium angles,
\begin{equation} \label{eq:impropers_terms}
 E_{\varphi} = \sum_{(X,A,B,C)} \begin{cases}
				k_\varphi^{XABC} \left( \varphi^{XABC} - \varphi_0^{XABC} \right)^2 & \text{if } \varphi_0^{XABC} < \tilde{\varphi} \\
				k_\varphi^{XABC} \left( \cos \varphi^{XABC}   - \cos \varphi_0^{XABC} \right)^2            & \text{otherwise} \quad .
		\end{cases}
\end{equation}
We decided on a threshold value $\tilde{\varphi}$ of 20$^\circ$ to consider a fragment planar, a choice that was guided by
the improper dihedral angles known from various compounds (recall the idealized values of $\varphi_0~=~0^{\circ}$ for planar and 
$\varphi_0~=~35.26^{\circ}$ for tetrahedral structures).

Non-covalent potentials are particularly challenging for a system-focused MM model, because it is hardly possible to parametrize these potentials from limited amounts of data of one equilibrium structure.
Therefore, universally parametrizable models for non-covalent contributions are needed.
We emphasize that the following simple approaches are by no means highly accurate for all of chemical space. However, we will later
show that they yield satisfactory results,
which can be improved on by our subsequent machine learning model.

The electrostatic term $E_{\text{estat}}$ is modeled by the sum over all Coulomb interactions between fixed atomic 
(partial) charges $q$ at the nuclear positions,
\begin{equation}
 E_{\text{estat}} =  \sum_{A\,<\,B} f^{AB}_{\text{estat}} \: \frac{q_A \, q_B}{r^{AB}} \quad .
\end{equation}
Here $f^{AB}_{\text{estat}}$ is a scaling factor:
for atoms $A$ and $B$ bonded to the same atom or to each other, $f^{AB}_{\text{estat}}$ is set to zero,
so that no electrostatic interaction is included as this has already been taken care of by the covalent terms,
and it takes a value of one in all other cases.
The background dielectric permittivity $\varepsilon$ is omitted in our model for the sake of simplicity and consistency.
It is typically set to unity in other atomistic models~\cite{riniker18}.

More advanced electrostatic models that activate multipole moments and intramolecular polarization have been discussed in the literature~\cite{rasmussen07, shi13},
but are not considered in our molecular mechanics model as we deliberately want to keep the force field simple
and correct for deficiencies in a subsequent machine learning step in a system focused manner. 

Van der Waals interactions are divided into an attractive (dispersion) and into a repulsive (Pauli repulsion) component.
Several correction schemes have been discussed~\cite{johnson09, grimme11_review, kriebel18} for
electronic structure methods which suffer from an insufficient description of dispersion interactions due to their incomplete description of electron correlation,
a common point of view considers the classical case of an attractive interaction resulting from 
oscillating dipoles originally introduced by London~\cite{london37},
which then lends itself to add attractive potential energy contributions defined in terms of {\it internuclear} distances
to the total electronic energy, analogously to the attractive part in Lennard-Jones potentials of classical force fields.

Here, we choose the London-type approach of the QMDFF force field for our model, which adopts 
functional form and parameters from the D3 semi-classical dispersion corrections for 
density functional theory (DFT)~\cite{grimme10}. First, this approach of distinctly definable atomic subentities within a molecule 
is consistent with the atomistic view of our resulting molecular mechanics model.
Second, the D3 scheme provides an extensively parametrized model that has been shown to yield accurate results within a DFT framework,\cite{goerigk11_1, goerigk11_2, risthaus13}
which can be considered a standard reference for our SFAM (see below). Moreover, D3
is readily applicable for the first 94 elements of the periodic table as parameters have been made available by the Grimme group.
The D3 expression for the dispersion energy is given by,
\begin{equation} \label{eq:dispersion}
 E_{\text{disp}} = - \sum_{A\,<\,B} f^{AB}_{\text{disp}} \left( \frac{C_6^{AB}}{(r^{AB})^6 + (f^{AB}_{\text{damp}})^6} + s_8 \: \frac{C_8^{AB}}{(r^{AB})^8 + (f^{AB}_{\text{damp}})^8}\right) \quad ,
\end{equation}
with the Becke-Johnson damping function~\cite{grimme11}
\begin{equation} \label{eq:BJ_damping}
  f^{AB}_{\text{damp}} = a_1R_0^{AB} + a_2 \quad .
\end{equation}

Obviously, the D3 model also includes an $r^{-8}$-term next to the standard $r^{-6}$-term known from the Lennard-Jones potential (and even higher
order terms were considered, but omitted for numerical stability reasons\cite{grimme11_review}). 
These higher-order terms can compensate for deficiencies in different DFT approximations.
Whether they can also be beneficial for the SFAM approach remains to be investigated.
As the global parameters $a_1$, $a_2$, and $s_8$ are unique to each functional,
we also reparametrize them for our force field (see section~\ref{sec:mm_parameter_optimization}).
The calculation of $C_6$ and $C_8$ coefficients, as well as the cutoff radii $R_0$, are taken from Grimme's D3 model~\cite{grimme10}.

In contrast to a typical DFT-D3 calculation, we do not re-evaluate the $C_6$ and $C_8$ coefficients for every structure, 
but instead treat them as fixed after the parametrization at the starting structure accelerating calculations significantly.
We found that this approximation introduces virtually no errors in our calculations
(note also that the predecessor of D3, i.e., D2, had these parameters fixed per atom pair).

The Pauli repulsion is modeled by an exponential potential adopting the pre-exponential factor from QMDFF,
\begin{equation} \label{eq:repulsion}
 E_{\text{rep}} = \sum_{A\,<\,B} f^{AB}_{\text{rep}} \: \frac{Z_{\text{eff},\,A} \: Z_{\text{eff},\,B}}{r^{AB}} \: \exp \left( \frac{-\beta r^{AB}}{R_0^{AB}} \right) \quad ,
\end{equation}
with the repulsion strength $\beta$ as a global fit parameter. The effective nuclear charges $Z_{\text{eff}}$ are also taken from QMDFF.
In analogy to the electrostatic scaling factor $f^{AB}_{\text{estat}}$, the scaling factors $f^{AB}_{\text{disp}}$ and $f^{AB}_{\text{rep}}$ are set to zero for atom pairs with a 1,2- or 1,3-bonding relationship.

An alternative choice for treating the van der Waals potentials was considered by adapting the functional form and parametrization of the 
universal force field (UFF)~\cite{rappe92},
which also provides a readily applicable model with parameters
for all elements. However, the D3 approach is more recent and offers some advantages, whereas UFF was found to be less accurate (see section~II.5). Grimme's D3 model allows us to easily reparametrize its global
parameters $a_1$, $a_2$, $s_8$, and $\beta$ in order to tailor the model to the other potential energy terms in SFAM without the need to reparametrize all element-specific parameters.
Moreover, we have already considered uncertainty quantification for the Grimme model~\cite{weymuth18} and how to improve it in a system-focused fashion~\cite{proppe19},
which straightforwardly extends to SFAM.

Furthermore, we implemented a slightly modified version of QMDFF's hydrogen bonding potential
to treat the formation of hydrogen bonds more accurately.
The following potential is applied for all of the entities D$-$H$\,\cdots\,$A (D, A = N, O, F, Cl) in the molecular system,
\begin{equation}
 E_{\text{hb}} =  -  \sum_{(D,H,A)} f^{DA}_{\text{damp},\, \text{hb}} \: \frac{c_{\text{hb}}^{DA}}{(r^{DA})^3} \quad ,
\end{equation}
with the damping function $f_{\text{damp},\, \text{hb}}$ composed of a distance and an angular component,
\begin{equation}
 f^{DA}_{\text{damp},\, \text{hb}} =    \left( 1 + \left( \frac{r_{\text{hb}}^{DA}}{\tilde{r}_{\text{hb}}}\right)^{12} \right)^{-1}  \left( \frac{1}{2} \left( \cos \phi_{\text{hb}}^{DHA} + 1 \right) \right)^6 \quad , \label{eq:hydrogen_bonded_interaction_damping}
\end{equation}
and the interaction strength $c_{\text{hb}}$ as a sum of the donor ($D$) and acceptor ($A$) contributions,
\begin{equation}
 c_{\text{hb}}^{DA} = k_{\text{hb},\,D} \: \frac{\exp \left( -\kappa_1q_D \right)}{\exp \left( -\kappa_1q_D \right) + \kappa_2} + k_{\text{hb},\,A} \: \frac{\exp \left( -\kappa_1 q_A \right)}{\exp \left( -\kappa_1 q_A \right) + \kappa_2} \quad . \label{eq:hydrogen_bonded_interaction_strength}
\end{equation}
In Eq.~(\ref{eq:hydrogen_bonded_interaction_damping}), $r^{DA}_{\text{hb}}$ represents the internuclear distance of the donor and the acceptor atoms, $\phi^{DHA}_{\text{hb}}$ is the angle between the donor, the hydrogen and the acceptor atoms, and
the constant damping parameter $\tilde{r}_{\text{hb}}$ is set to 4~\AA.
The parameters $k_{\text{hb},\,D}$ and $k_{\text{hb},\,A}$ of Eq.~(\ref{eq:hydrogen_bonded_interaction_strength}) are globally fitted (see section~\ref{sec:mm_parameter_optimization}) and specific to any of the possible elements involved,
$q_D$ and $q_A$ are the atomic (partial) charges of the donor and acceptor, respectively, and the values of $\kappa_1$ and $\kappa_2$ are taken from QMDFF~\cite{grimme14} ($\kappa_1 = 10$, $\kappa_2 = 5$).
The choice of the latter is not critical as they merely regulate the decrease of the interaction strength with respect to the increase of the atomic charge on the donor or acceptor atom.

\subsection{Atom-type definitions}
\label{sec:atom_type_definitions}

Classical force fields typically rely on the definition of atom types.~\cite{riniker18}
Atom types are categories into which the atoms of a system are assigned based on some criteria such as their element type and their chemical environment,
e.g., the functional group they are part of.
In the resulting molecular mechanics model, atoms of the same atom type are treated together as one entity in the definition of parameters.
Force fields such as AMBER~\cite{cornell95} and CHARMM~\cite{brooks83} predefine a set of atom types on which their parametrization is based,
hence limiting their applicability to arbitrary systems.

In a system-focused parametrization framework, a definition of atom types is not necessary, but provides two 
advantages and is therefore employed in this work.
First, it implicitly implements local symmetry constraints on the model parametrization.
For instance, treating the three hydrogen atoms of a methyl group individually would result in a model with three different C$-$H bond force constants
arising from asymmetries in the static arrangement of the chemical environment in the reference structure.
Moreover, we identified numerical instabilities in the fitting procedure related to the absence of such symmetry constraints.
Second, the partial Hessian fit to be described in section~\ref{sec:mm_parameter_optimization} benefits from more reference data to fit a specific parameter,
which can be provided by grouping very similar potential contributions according to the atom types involved.
As a consequence, there are fewer parameters to fit in total so that the parametrization process is accelerated by choosing a coarser atom-type assignment.

As these atom types will always be arbitrarily defined to some extent and they are neither necessary in quantum chemical calculations nor in machine learning approaches
of force fields~\cite{botu16, bartok17, glielmo17, chmiela18},
we decided on a flexible implementation of atom-type levels systematically increasing the ``uniqueness'' of the model parameters in order to limit arbitrariness while
exploiting the observed benefits during the parametrization procedure.
We start by defining atom types solely on an element-type basis,
subsequently adding the number of covalently bonded neighbors and their element types, and finally adding the information of more layers of bonded atoms.
In all results presented in this work, we chose a fine-grained atom-type definition, i.e.,
combining all atoms of the same element and the same set of neighboring elements into one atom type.
For hydrogen atoms, we combine all atoms that are bonded to the same atom type into one type of hydrogen atoms.
Although we found no issues with this atom-type definition so far, we emphasize that is does not cover all features that might be desired. For instance, aromatic and alkene carbon atoms are currently
not distinguished. Our software implementing SFAM allows for an easy extension of the atom-type definition that then incorporates more atom types if deemed convenient.

Dihedral angle parameters are typically characterized by the four atom types involved in its topological definition.~\cite{riniker18}
After a careful study of several options in combination with our parametrization and atom-type classification scheme, we decided on a two-atom based dihedral angle definition instead:
the atom types of the two central atoms of a dihedral angle determine the parameters in the expression for the potential.

In this work, we do not employ atom types to define the non-covalent potential energy functions, but instead treat every atom individually when calculating the non-covalent interactions,
i.e., each atom is assigned a specific partial charge from a quantum chemical calculation and
each atom pair has its own set of dispersion coefficients. This makes SFAM's non-covalent potentials more accurate and we have not
observed any issues with this strategy so far.

\subsection{Optimization of MM parameters}
\label{sec:mm_parameter_optimization}

The aim of our automated parametrization procedure is to generate a satisfactory base model for subsequent on-the-fly improvement
from local quantum chemical information on the PES. 
To begin the parametrization, the reference molecular structure and its Hessian matrix must be obtained from an electronic-structure
calculation. In principle, this can be done at an arbitrary level of approximation, i.e.,
ranging from semi-empirical to DFT to explicitly correlated coupled cluster calculations.
Our model parameters fall into three categories for parametrization, namely those 
assigned based on the system's reference structure, parameters fitted to the Hessian matrix, and globally fitted parameters.

First, we assign the equilibrium values $r_0$, $\alpha_0$, and $\varphi_0$ of Eqs.~(\ref{eq:bond_term}),~(\ref{eq:angle_term})~and~(\ref{eq:impropers_terms}) corresponding to the mean value of their occurrences in the reference structure.
This structure is a local electronic energy minimum on the PES that can be chosen to closely resemble the starting point of a given configuration space exploration,
e.g., an MD trajectory or an automated reaction network exploration~\cite{simm17, simm18, unsleber19}.
For equilibrium improper dihedral angles below the planarity threshold $\tilde{\varphi}$ introduced in section~\ref{sec:model_definition}, we set its value to become $\varphi_0 = 0^{\circ}$.
Furthermore, the parameters $n$ and $\theta_0$ of the dihedral potentials for a group of sequentially bonded atoms $(A,B,C,D)$ are assigned
based on rules adapted from the work of Van Speybroeck and coworkers~\cite{vanduyfhuys15}.
The periodicity $n$ is determined by
\begin{equation}
 n = \frac{(N - 1) (M - 1)}{\mathrm{gcd}(N - 1,\,M - 1)} \quad , \label{eq:periodicity}
\end{equation}
where $N$ and $M$ are the number of atoms bonded to the atoms $B$ and $C$, respectively,
and $\mathrm{gcd}(k,\,l)$ represents the greatest common divisor of the integers $k$ and $l$.
To determine the phase shift for a given dihedral potential, the image $I$ of the dihedral angle $\theta$,
\begin{equation}
 I = \left| \theta \right| \ \mathrm{mod} \ \frac{2\pi}{n} \quad ,
\end{equation}
is assigned to one of the three intervals $\left[0, \frac{\pi}{3n}\right]$, $\left[\frac{2\pi}{3n}, \frac{4\pi}{3n}\right]$ and $\left[\frac{5\pi}{3n}, \frac{2\pi}{n}\right]$.
This procedure is repeated for all dihedral angles that correspond to this potential, i.e., the rotation around the bond of atom $B$ and $C$.
If $I$ can be assigned to the same interval for all of these dihedral angles, the phase shift can be set to $\theta_0 = n\bar{I}$, where $\bar{I}$ is the mean of all $I$ for this dihedral angle.
When this is not possible, the dihedral potential cannot be described well by expression~(\ref{eq:dihedral_potential}) and is not included in the model.
More complex potential energy functions may be developed for these cases in future work.
Examples for the assignment of phase shifts for dihedral potentials can be found in the Supporting Information of the original work by Van Speybroeck and coworkers.~\cite{vanduyfhuys15}.

The fixed atomic charges in the electrostatic part of our model are also assigned based on the reference structure.
In the literature, several charge models have been evaluated~\cite{verstraelen12, vilseck14}. In agreement with these studies,
we found schemes based on corrections of the Hirshfeld partitioning~\cite{hirshfeld77} of the electron density,
namely the iterative Hirshfeld scheme~\cite{bultinck07} and the Charge~Models~1-5~\cite{marenich12}, to be most suitable
for describing intermolecular interactions within our MM model.
We chose the Charge~Model~5 (CM5) owing to its availability in the Gaussian 09 software package~\cite{gaussian09} 
for all atomic charges in this work and defer an investigation of
possible advantages of the iterative Hirshfeld scheme over the CM5 approach to future work.
Furthermore, as the atomic charges strongly depend on the chemical environment, e.g., a solvent,~\cite{mei15, riquelme18}
this environment should be described in the reference calculations as accurately as possible, for instance, by
including solvent molecules explicitly (i) through an automated microsolvation procedure such as the one that we described in Ref.~\cite{simm19})
that can be adapted to our fragmentation scheme for large molecular structures or (ii) by treating these molecules implicitly in a polarizable continuum model.

Second, the force constants of our model potentials are fitted to local curvature information of the potential energy obtained from the Hessian matrix of the system.
To obtain an optimal parameter vector $\mathbf{p}$ composed of the parameters $k_r$, $k_\alpha$, $V_\theta$,
and $k_\varphi$, a multi-step optimization procedure is employed,
which is based on the partial Hessian fitting scheme proposed by Hirao and coworkers~\cite{wang16, wang18}.
In each step, only one parameter $p_i$ is optimized, while the others are kept constant.
The optimization's loss function is constructed based on partial Hessian submatrices corresponding to atom pairs in a set $\mathcal{P}$,
which is composed of different atom pairs in each step.

The partial Hessian matrix $\mathbf{H}_{AB}$ for an atom pair $( A,\,B )$,
\renewcommand\arraystretch{2.0}
\begin{equation}
\mathbf{H}_{AB} = 
  \begin{pmatrix}
\frac{\displaystyle\partial^2 E_{el}}{\displaystyle\partial x_A \partial x_B} & \frac{\displaystyle\partial^2 E_{el}}{\displaystyle\partial x_A \partial y_B} & \frac{\displaystyle\partial^2 E_{el}}{\displaystyle\partial x_A \partial z_B} \\
\frac{\displaystyle\partial^2 E_{el}}{\displaystyle\partial y_A \partial x_B} & \frac{\displaystyle\partial^2 E_{el}}{\displaystyle\partial y_A \partial y_B} & \frac{\displaystyle\partial^2 E_{el}}{\displaystyle\partial y_A \partial z_B}  \\
\frac{\displaystyle\partial^2 E_{el}}{\displaystyle\partial z_A \partial x_B} & \frac{\displaystyle\partial^2 E_{el}}{\displaystyle\partial z_A \partial y_B} & \frac{\displaystyle\partial^2 E_{el}}{\displaystyle\partial z_A \partial z_B}  \\
\end{pmatrix}
\quad .
\end{equation}
is a $(3\times3)$ submatrix of the total Hessian matrix of the system ($E_{el}$ is the system's total electronic energy, not to be confused with the electrostatic energy $E_{\text{estat}}$ in the MM model).
The optimized force field parameters are obtained by a least squares minimization, where the mean squared error (MSE)
of the elements of the partial force field Hessian $\mathbf{H}^{\text{MM}}$ with respect to its reference counterpart $\mathbf{H}^{\text{ref}}$ can be expressed
as the squared Frobenius norm of the difference matrix,
\renewcommand\arraystretch{1.5}
\begin{equation}
 \text{MSE}_{AB}\!\left( p_i \right) = \lvert \, \mathbf{H}_{AB}^{\text{ref}} - \mathbf{H}_{AB}^{\text{MM}}\!\left(p_i\right) \, \rvert ^2 \quad .
\end{equation}
For all atom pairs $( A,\,B )$ in the set of relevant pairs $\mathcal{P}$, the sum of all MSEs is minimized with respect to the force field parameter $p_i$,
\begin{equation}
 \label{eq:parameter_opt}
 p_{i,\,\text{opt}} = \underset{p_i}{\text{arg min}} \left\lbrace \sum_{( A,\,B ) \,\in\, \mathcal{P}} \text{MSE}_{AB}\!\left( p_i \right) \right\rbrace \quad .
\end{equation}
The Levenberg-Marquardt algorithm~\cite{levenberg44, marquardt63} can then be applied to solve Eq.~(\ref{eq:parameter_opt}) efficiently.

In the first step of the optimization procedure, the half barrier heights $V_\theta$ of the dihedral potentials are optimized.
The choice of a parameter $V_\theta$ affects all submatrices $\mathbf{H}_{AB}^{\text{ref}}$, with $A$ and $B$ being any of the four atoms defining dihedral angles with that specific half barrier height $V_\theta$.
We construct our set $\mathcal{P}$ as all of these atom pairs, which are in a 1,4-bonding relationship, because the corresponding submatrices $\mathbf{H}_{AB}^{\text{ref}}$ only depend on the parameter $V_\theta$.
The remaining parameters are kept constant at their initial values during the optimization. Once optimal half barrier heights $V_\theta$ are determined, the angle force constants $k_\alpha$ are optimized next.
Here, $\mathcal{P}$ consists of all atom pairs in a 1,3-bonding relationship, that correspond to a given force constant $k_\alpha$.
Analogously, the bond force constants $k_r$ are optimized subsequently based on atom pairs in a 1,2-relationship.
This procedure guarantees that each parameter is optimized with respect to data, which only depends on the parameter itself or on previously optimized ones.
Choosing a different order of parameter optimization would produce a bond parameter being explicitly dependent on a yet unoptimized angle or dihedral parameter, since
the partial Hessian $\mathbf{H}_{AB}^{\text{MM}}$ of two atoms A and B bonded to each other also depends on all the parameters for the angles and dihedrals involved.
For a more detailed discussion of the chosen procedure, the reader is referred to the original work by Hirao and coworkers~\cite{wang16}.

For the rarely appearing improper dihedral force constants $k_\varphi$, we deviate from this scheme. The parameters $k_\varphi$ are optimized after all the others.
We found, that placing all atom pairs $( A,\,B )$ in the set $\mathcal{P}$,
where $A$ is the central atom and $B$ corresponds to any of the four atoms related to the improper dihedral angle including $A$ itself, yielded satisfying results.
By this last step of optimizing improper dihedral force constants, we introduce an error into our previously optimized parameters, 
because they may have depended on the unoptimized improper dihedral parameters.
However, this strategy proved to perform well and may be improved on in future work.

We emphasize that the choice of a partial Hessian fitting approach relying only on local information is highly advantageous for the purpose of parametrizing large systems,
for which the evaluation of an entire Hessian is computationally not feasible.
Since a force field parameter is only fitted to specific blocks of the Hessian matrix,
the total Hessian may be approximated by partially constructing it from Hessians of molecular fragments and still be employed as the input for our method.
Furthermore, this optimization is very efficient, so that the bottleneck of the parametrization is the quantum chemical reference calculation,
especially the computation of the Hessian matrix.

The non-covalent interaction terms $E_{\text{nb}}$ introduced in section~\ref{sec:model_definition} contain parameters, that are not system-focused, but predetermined.
For some of these parameters, a system-focused parametrization will be considered for future work.
In general, parametrizing the non-covalent terms in a system-focused manner from reference data of a single local minimum energy structure is more difficult than for covalent terms, therefore
we propose to treat the arising deficiencies in the MM model in our machine learning approach to be described in part~\ref{sec:ML} of this work.
The parameters $a_1$, $a_2$, $s_8$, and $\beta$ in Eqs.~(\ref{eq:dispersion})-(\ref{eq:repulsion}) control the van der Waals interactions of our model.
A fit to the interaction energies and dissociation curves of 22 molecular dimers yielded optimal values of $a_1=0.1$, $a_2=7.1$~bohr, $s_8=4.6$ and $\beta=7.4$.
The element dependent interaction strengths $k_\text{hb}$ introduced in Eq.~(\ref{eq:hydrogen_bonded_interaction_strength}) were fitted to the interaction energies of small hydrogen-bonded dimers.
For the molecules considered in this work, they were found to be $k_{\text{hb}}$(N)~=~0.6~a.u., $k_{\text{hb}}$(O)~=~0.7~a.u., $k_{\text{hb}}$(F)~=~3.2~a.u. and $k_{\text{hb}}$(Cl)~=~4.2~a.u., respectively.
Details about the fit of these global parameters can be found in the Supporting Information.

Weak dispersion interactions can be expected to hardly affect the second derivatives in the partial Hessian matrices and the atomic charges.
Accordingly, we did not find that, for instance, including dispersion corrections for a DFT reference method produces a 
noteworthy effect on the optimized SFAM force constants.
However, as dispersion interactions in DFT can be switched either on or off during the parametrization step,
we note that for a consistent parametrization 
the same choice should then be applied to the SFAM parameter optimization process. 
If dispersion interactions cannot be switched off in the reference calulations (e.g., in the case of a coupled cluster model chosen as reference),
the dispersion interactions will be switched on for the SFAM parametrization of force constants.

The typical MM model dissection of the electrostatic interactions which are unambiguously defined only for the elementary particles (electrons, nuclei) that constitute the molecules,
produces partial atomic charges and spring models
decoupling bonded and non-bonded interactions which are exact only for the reference structure for which the parametrization was carried out.
Currently, this deficiency in our model is corrected in the subsequent ML model improvement step.
However, since such a decoupling may introduce errors in the kcal/mol range~\cite{konig15, hudson19},
further analysis in future work should be beneficial.

\subsection{Fragmentation of large systems for parametrization}
\label{sec:subsystem_parametrization}

If the evaluation of the entire system's Hessian is not feasible due to its size, the MM model must be generated by a divide and conquer approach.
Therefore, quantum chemical reference calculations generally need to be performed on molecular fragments of the full system.
Such fragments will often be valence-saturated by hydrogen atoms if the fragmentation protocol cuts through bonds, which is, however, not easily possible for any arbitrary system.
This problem can be alleviated by the application of wavefunction-based embedding methods~\cite{manby12, fornace15, muhlbach18, lee19}
which ensure that the system can be arbitrarily partitioned. In the present proof of principle study, we implemented the standard approach of
valence saturation by hydrogen atoms first.

\begin{figure}[H]
\begin{center}
\vspace*{-0.5cm}
\includegraphics[width=0.9\textwidth,trim={0cm -1.2cm 0cm -0.8cm},clip]{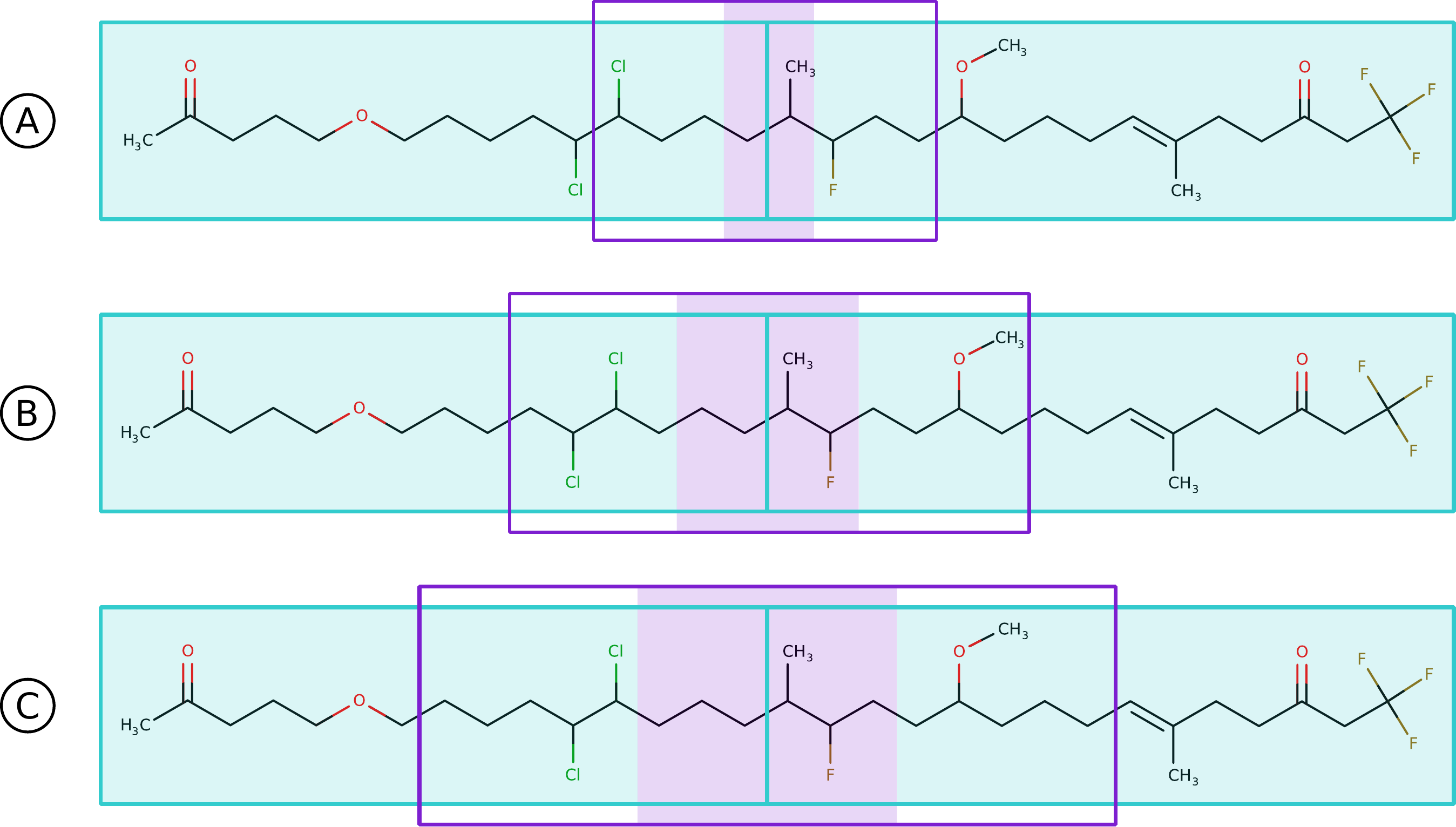}
%trim={<left> <lower> <right> <upper>
\end{center}
\vspace*{-0.8cm}
\caption{\label{fig:subsystem_parametrization}\small Three different partitionings of the
pseudo-one-dimensional molecular structure OCE for the subsystem parametrization procedure.
The rectangular boxes define the three subsystems (i.e., the line of a box embraces all atoms contributing to the subsystem indicated by that box)
for which Hessian matrices are calculated separately. The resulting dangling bonds of these subsystems are saturated by hydrogen atoms.
All partial Hessian submatrices corresponding to atoms only present in the cyan shaded regions are taken from the Hessian matrices calculated
for the subsystems defined by the two cyan boxes, i.e., by the left and right subsystems.
All submatrices including at least one of the atoms in the purple shaded region 
are taken from the central subsystem's Hessian matrix, i.e., from that of the subsystem defined by the purple rectangular box.
The size of the buffering regions increases from A to C to highlight that this parameter can be controlled.}
\end{figure}

In the following, we demonstrate the concept of the fragment based MM model parametrization with two examples. The first one is an
organic chain-like ether~(OCE) with several functional groups depicted in Fig.~\ref{fig:subsystem_parametrization} that allows us to
describe the fragmentation principles and reference-data choice from the fragments. The second one, the protein plastocyanin~\cite{redinbo94},
presents a three dimensional structure of sufficient size to highlight fragmentation in three dimensions at a small nanoscale
example.

The molecular structure of OCE was prepared in a linear conformation and then optimized with the standard DFT model
PBE-D3/def2-SVP. The structure optimization locked into the linear structure (i.e., it did not coil up) as desired for our purpose.
We apply a subsystem-based parametrization scheme with a molecular partitioning strategy similar to the GMFCC approach by Zhang and coworkers~\cite{zhang03, he06, wang13}.
The system is divided into two halves at a bond close to its center and the ends of the resulting subsystems are saturated with hydrogen atoms
by placing them along the cut bond vector at a distance determined by the van der Waals radii of the involved elements.

It is obvious, that the partial Hessian matrices evaluated for atoms close to the separating cut or to two atoms that were separated into the different subsystems cannot be employed to fit the MM parameters.
Therefore, a third subsystem is defined enclosing the bond that was cut in the first step. In general, $N+(N-1)$ Hessians are therefore calculated for a division
of a pseudo-linear system into $N$ subsystems employing this particular fragmentation strategy. We include buffering regions for all subsystems located between the relevant atoms for the fit and the border of the subsystem.
Furthermore, we compare differently sized buffering regions to study their effects on the errors of the force constants
introduced by a subsystem parametrization in contrast to the typical approach based on the entire Hessian matrix.

Table~\ref{tab:subsystem_parametrization} gives an overview of the deviations for the parametrized force constants introduced by our subsystem parametrization approach
for the three different partitionings of the OCE molecule. From the results presented in that table it is clear that a large enough choice (B or C) of the buffering region
leads to very small errors in the force constants of always less than 1.1 percent. While the mean deviations are also acceptable for the partitioning A,
because distant atoms are not affected by the small buffering region, errors larger than 50 percent are observed for some force constants.
Hence, reasonably sized buffering regions are important to obtain accurate force constants even in the subsystem-based scheme.
In the results presented, the subsystems were not reoptimized after the separation, as we found 
that such a reoptimization only introduces slightly larger errors in the obtained force constants.

\begin{table}[H]
\renewcommand{\baselinestretch}{1.0}
\renewcommand{\arraystretch}{1.0}
\caption{\label{tab:subsystem_parametrization}\small Mean and maximum relative deviations for the SFAM force constants 
of the OCE molecule obtained from
subsystem based parametrization in comparison with results from a parametrization based on the full-system Hessian.
In both cases, the reference method for the parametrization was PBE-D3/def2-SVP. All data are given in percent.
The three different partitionings A, B and C, shown in Fig.~\ref{fig:subsystem_parametrization}, are compared.
}
\begin{center}
\begin{tabular}{l l c c c} 
\hline
\hline
 & & A & B & C \\
\hline 
\multirow{2}{*}{$k_r$} & mean & 2.98 & 0.02 & 0.01 \\
& max. & 58.2 & 0.18 & 0.18 \\
\hline
\multirow{2}{*}{$k_{\alpha}$} & mean & 1.10 & 0.04 & 0.02 \\
& max. & 65.6 & 0.29 & 0.28 \\
\hline
\multirow{2}{*}{$V_{\theta}$} & mean & 2.33 & 0.22 & 0.07 \\
& max. & 23.1 & 1.08 & 0.28 \\
\hline
\hline
\end{tabular}
\renewcommand{\baselinestretch}{1.0}
\renewcommand{\arraystretch}{1.0}
\end{center}
\end{table}

For an arbitrary three-dimensional system, however, two issues arise.
First, since the atoms of a typical molecular system are not uniformly distributed in space, it is a non-trivial task to define a minimum number of cubes or spheres such that any structure is separated into subsystems
of approximately equal number of atoms without most fragments consisting of just a few atoms while others comprise a very large amount.
We emphasize that for a protein the aforementioned strategy could be applied to its primary structure, but in general this cannot be safely assumed.
Second, the number of fragments additionally introduced for buffering regions would also increase significantly
for large three-dimensional systems.

We therefore prefer the latter strategy of creating one spherical molecular fragment
centered around each atom. This approach offers the advantage of (i) a trivial implementation, (ii) a strictly linear scaling of the number of reference calculations with the
number of atoms, and (iii) a redundancy of reference data for each parameter, which can be exploited for an additional uncertainty assessment of the base MM model and to compensate
missing reference data resulting from failed electronic structure calculations for specific fragments.

The radius of the sphere defining the fragments must be chosen in a way that the atoms in a 1,4-bonding relationship with the central atom are included with a sufficient buffering region at the boundary of the sphere.
We found that a radius of 5-7~\AA~is a reasonable choice. For some systems, a radius of 7~\AA~will be feasible, however, for others it may be chosen smaller for all fragments in order
to be of computationally treatable size. Note that the latter strongly depends on the choice of the electronic structure method applied for the generation of reference data.
Additional conditions must be satisfied during the fragmentation process. First, the resulting fragment structure needs to be chemically reasonable for the quantum chemical reference calculations
to converge and to yield reliable results (at least as long as no embedding approach is applied that can cut through covalent bonds without significant loss of accuracy).
This requires the partitioning of the structure to be controlled in a way that allows for straightforward saturation of the cleft bonds.
The procedure additionally requires the knowledge of charges and unpaired electrons at localized sites in the structure such that each fragment can be assigned a reasonable total charge and spin multiplicity.
Clearly, this assignment must be automated and may require more than one reference calculation to decide on a particular electronic structure for the fragment in an automated fashion.
Second, during the structure optimization of each molecular fragment, which, in general, will consist of more than one molecule, structural constraints are necessary to ensure that the structure
will not relax into an energy minimum that would not be accessible in the full system due to the backbone structure of the molecular framework.

Considering all of the aforementioned requirements, we implemented the following fragmentation procedure.
We begin by generating a fragment from a spherical cutout around each atom while keeping track of the elements and the connectivity of the atoms involved in cleft bonds.
If the cleft bond cannot be straightforwardly saturated, we will recursively follow the bonds encoded in the system's molecular graph representation until a cleavable bond is located.
Hence, atoms more distant from the fragment's central atom than the chosen radius are also included in the fragment to ensure obtaining chemically reasonable fragments.
This algorithm requires the covalent bonds of the system to be known prior to the reference calculations.
We therefore apply a covalent radii based connectivity guess, which is validated and possible corrected once the first-principles derived bond order information becomes available.
As one molecular fragment may consist of more than one molecule (molecular subgraph),
subgraphs that consist of a small number of atoms will be discarded.
For example, a lonely hydroxy group from a disconnected part of the full system extending into the fragment sphere will be ignored
if it is close to the outer boundary of the sphere.
The cleft bonds are saturated by hydrogen atoms, which are placed along the previous bond vector at a distance determined by the covalent radii of the involved elements.
Finally, we enforce a minimum number of atoms per fragment to ensure that the chemical environment around a given atom is reasonably well represented in all cases. This is achieved by
repeating the process described above with a larger radius for fragments of insufficient size.

For the subsequent quantum chemical structure optimization of the (saturated) molecular fragment,
we constrain a maximum number of two bonds per molecule by keeping the Cartesian coordinates of the involved atoms fixed during
the optimization. These two bonds are chosen from the set of all saturated bonds by determining the bond most distant from the central atom and 
the one with the largest graph distance to it (i.e., largest number of bonds between them), also enforcing a minimum graph distance of four.
If such a second bond does not exist, because only one bond was saturated inside the given molecule or all pairs of bonds have a smaller graph distance than the enforced minimum,
we will constrain only the first bond for that specific molecule.
By careful testing of this procedure against alternatives (such as constraining all saturated bonds)
it was found to be a good compromise between assuring the system's flexibility during the optimization and, at the same time,
approximately preserving the spatial arrangement of the molecules inside the fragment.

For each optimized fragment, the Hessian matrix, CM5 charges, and the covalent bond orders are calculated.
The reference data needed during the parametrization procedure discussed in section~\ref{sec:mm_parameter_optimization} is assembled from these fragment calculations in the following way.
The $(3\times3)$ partial Hessian matrix for the atoms $A$ and $B$ is taken from the full Hessian calculated for the fragment, which has atom $A$ at its center.
If this Hessian is not available because of a failed calculation, the fragment with $B$ at its center will be exploited.
Furthermore, if neither Hessian is available, the fragments with the covalent neighbors of $A$ and $B$ at their centers will be employed instead.
An analogous procedure is applied to obtain equilibrium molecular structure data
(e.g., the optimal bond length between $A$ and $B$), the atomic charges, and the first-principles derived connectivity.

\begin{figure}[H]
\vspace*{-0.5cm}
\begin{center}
 \includegraphics[width=0.8\textwidth]{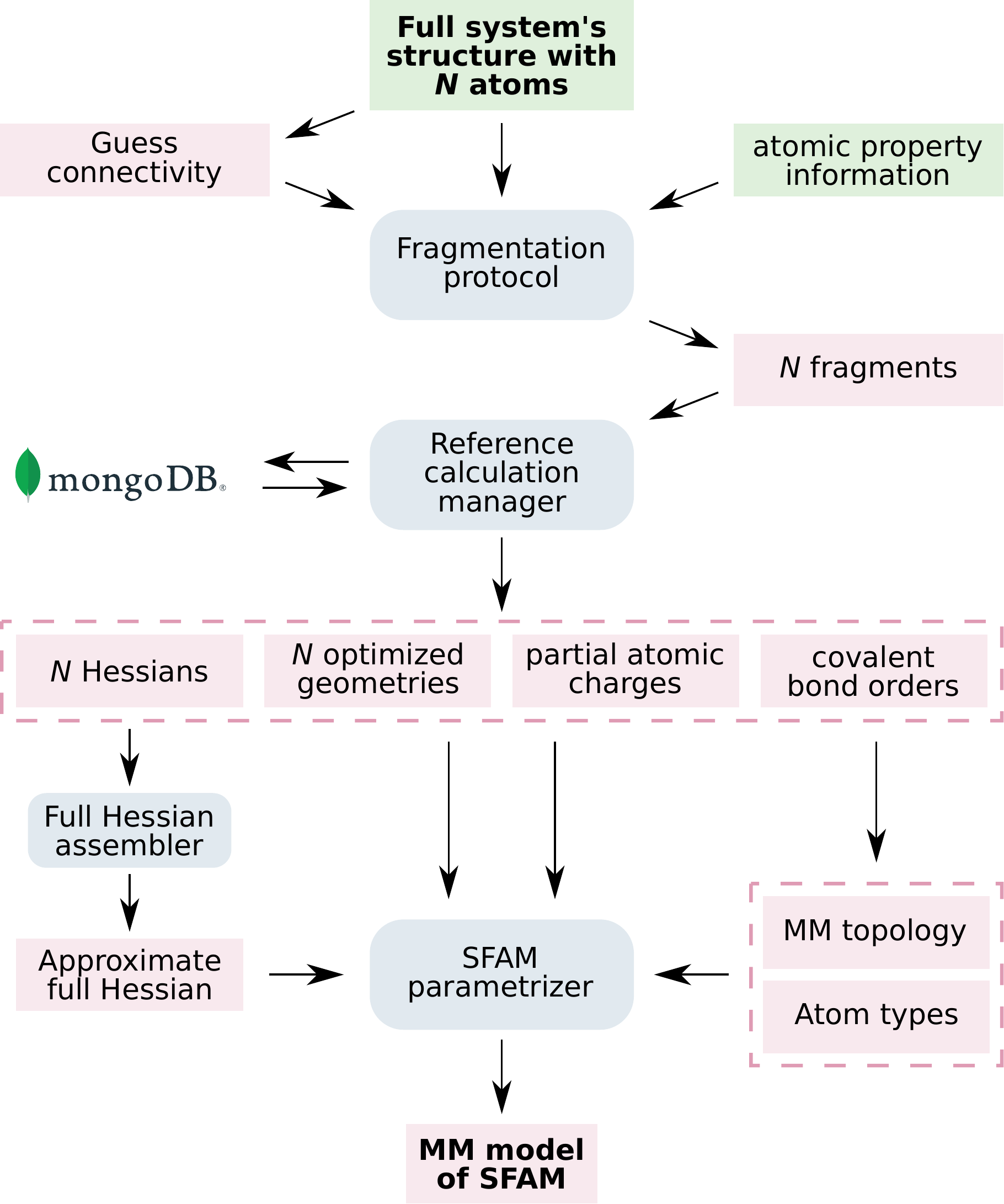}
 \caption{\label{fig:overview_parametrization}\small Overview of the SFAM parametrization procedure for large systems that require fragmentation.
 The red boxes represent the data objects of the parametrization, while the essential
 algorithmic subunits of the procedure are presented in blue boxes. The input data is shown in green boxes. Here, we present the essential elements of our algorithm, starting
 with the fragmentation process, which yields the molecular fragments for which reference data are calculated. Based on these reference data, the MM model is obtained
 from the SFAM parametrizer. Note that the approximate full Hessian for the nanoscale system assembled from the fragment Hessians only contains the $(3\times3)$ partial Hessian
 (sub-)matrices needed in the parametrization procedure and is therefore sparse.}
  \end{center}
  \vspace*{-0.5cm}
\end{figure}

We stress that the large amount of reference calculations can be trivially parallelized and the computing time would be independent of
fragment number provided that as many computing cores are available as there are fragments. For instance, a system with 10\,000
atoms separated into fragments ranging from 20 to 150 atoms (which is reasonable
for the radii discussed above) and a high performance computing infrastructure with
500 cores results in 20 quantum chemical reference calculations for each core, typically finished
in a few hours to days on modern computing hardware.

These reference calculations are managed through a \texttt{MongoDB} database~\cite{mongodb}, which we previously employed for our automated
reaction network exploration software~\cite{simm17, simm18, unsleber19}.
In this set-up, our parametrization software writes all information about the reference calculations into the database, which is then processed by a second program, which deposits the results
of the completed calculations back into the database for subsequent analysis by the parametrization algorithm.
This procedure allows for efficient data management and the implementation of the trivial parallelization mentioned above.

\begin{figure}[H]
\begin{center}
 \includegraphics[width=0.9\textwidth]{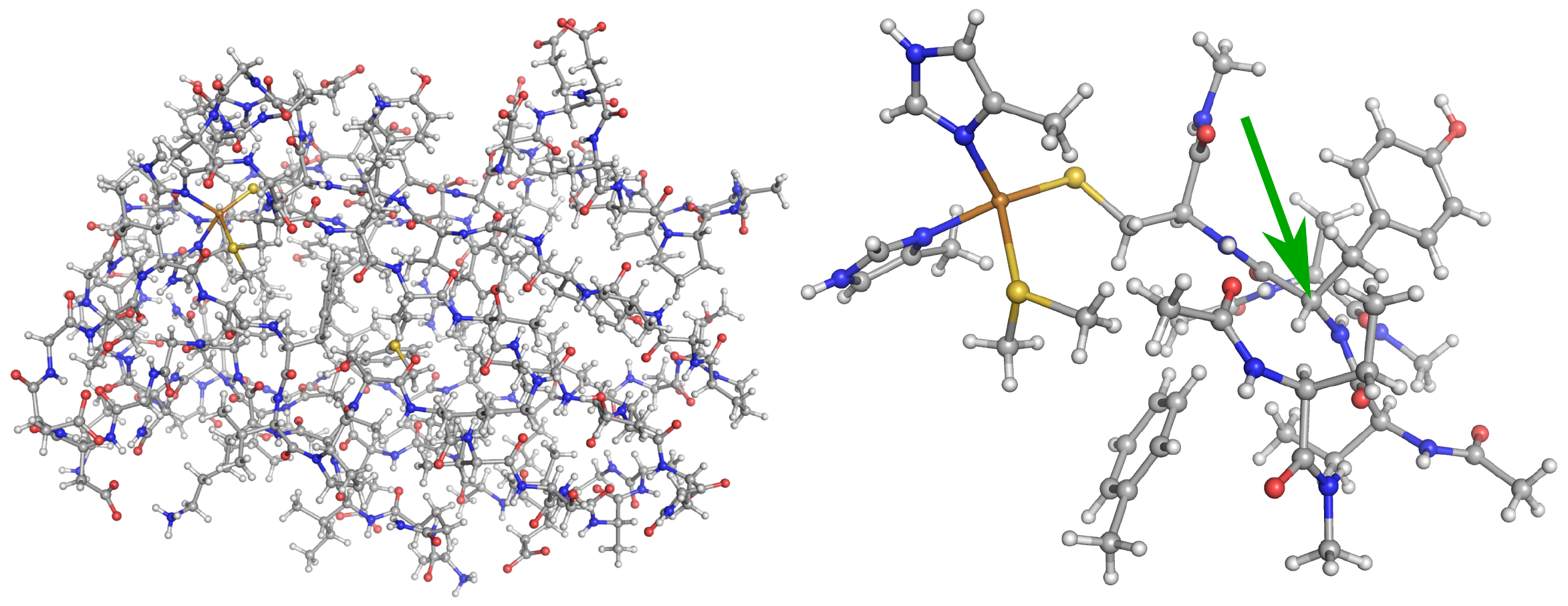}
 \caption{\label{fig:plastocyanin_structures}\small The molecular structures of plastocyanin (left) and its largest fragment with 152 atoms (right) obtained during the SFAM fragmentation procedure.
 The green arrow points to the central atom of the fragment around which it was constructed.}
   \end{center}
   \vspace*{-0.5cm}
\end{figure}

A scheme summarizing the workflow of the parametrization of nanoscale systems by the strategy discussed in this section is shown in Fig.~\ref{fig:overview_parametrization}.
We now proceed to study this strategy at the second example, the copper-containing protein plastocyanin~\cite{redinbo94} (PDB ID: 1IUZ) with 1412 atoms.
Plastocyanin is too large for a single reference Hessian calculation on the entire system in a reasonable time.
Therefore, the fragmentation based parametrization procedure is necessary. We applied the strategy discussed above with a radius of 5.5~\AA. During this process,
only those bonds were considered cleavable that were of the form $\text{C}_{\text{sp}^3}-\text{C}$ and $\text{C}_{\text{sp}^3}-\text{N}$, i.e., the atom inside the fragment
being an sp$^3$-carbon atom and the atom outside being either a carbon or a nitrogen atom. A minimum fragment size of 20 atoms was enforced.
The information about localized charges, e.g., in protonated amino groups, as well as unpaired electrons at the copper atom, were manually provided as an input to the fragmentation algorithm.
However, we emphasize that this information can, in principle, be generated for a given system in an automated fashion
by a few additional quantum chemical calculations that decide on one of the local states.
Sizes between 20 and 152 atoms were obtained for the resulting molecular fragments. The mean fragment size was 73.5 atoms.
The structure of plastocyanin together with the structure of the largest fragment is depicted in Fig.~\ref{fig:plastocyanin_structures}.
Moreover, a histogram displaying the distribution of fragment sizes is provided in the Supporting Information.

For the parametrization, the fragments were optimized under the aforementioned Cartesian constraints with the RI-PBE-D3/def2-SVP~\cite{whitten73, dunlap79, vahtras93, perdew96, grimme10, weigend05}
combination of density functional and basis set. The Hessian matrix for each fragment was calculated applying the same method. CM5 charges and covalent bond orders were obtained with PBE/def2-SVP.
The atomic charges were calculated with the \texttt{Gaussian09} software package~\cite{gaussian09}, while
all other electronic structure calculations were carried out with the \texttt{ORCA 4.2} program~\cite{neese12, neese18}.
Two atoms were considered bonded when their Mayer bond order was larger than 0.4, however, for one expected bond between a copper and a sulfur atom a bond order of only 0.31 was obtained.
Therefore, we added this bond manually. Obviously, this is not acceptable in a fully automated procedure.
However, it is easy to overcome as one can either implement more advanced, automated and reliable first-principles derived bond assignment methods~\cite{glendening12, knizia13, manz17}
or extend SFAM to comprise bond formation and breaking processes similar to a reactive force field~\cite{vanduin01, senftle16}.

Applying the atom-type definitions described in section~\ref{sec:atom_type_definitions}, we obtained 45 different atom types for plastocyanin, which results in 284 force constant parameters to be optimized.
To demonstrate that the automatically parametrized
molecular mechanics model obtained for this large system can be straightforwardly
applied, we performed an 800~ps MD simulation of the isolated molecule (in vacuum
and without peridodic boundary conditions) at a constant temperature of 300 K
adjusted by a Berendsen thermostat~\cite{berendsen84}.
The results of this simulation are discussed in the Supporting Information.

Despite the fact that this strategy was successfully applied to parametrize plastocyanin, we made a few observations that deserve mentioning.
First, in two cases proton transfer reactions involving NH$_3^+$ and CO$_2^-$ groups were observed during the structure optimizations of the fragments.
This results in the partial Hessian matrix $\mathbf{H}\left(\text{H}^{\text{trans}}, \text{N} \right)$ and the equilibrium bond distance $r_0\left(\text{H}^{\text{trans}}, \text{N} \right)$
being inaccurately calculated for this atom pair. Furthermore, the observed process also affects the reference data obtained for the carboxylate anion.
Note, however, that this is not an issue of the SFAM fragmentation protocol, instead it is a general problem of the reference calculations that would also occur for a structure optimization of the full system.
Since in our example we encountered this proton transfer reaction for only a small fraction of the NH$_3^+$ groups and the parameters are fitted
to all of the reference data gathered for these atom types, our results were hardly affected by this problem.
However, this will not be the case in general and therefore the possibility of unwanted chemical reactions occurring during fragment optimizations can be monitored in every parametrization
and eventually prevented by constraining the structure optimizations to prevent these processes.

A second issue arises from the fact that only a limited amount of reference calculation failures resulting from, for instance, SCF convergence problems, can be compensated
by exploiting the data redundancy obtained from the atom-centered spheres fragmentation approach. A fully automated parametrization procedure demands to prevent manual interference
during the parametrization process.
One can either choose SCF convergence control settings that ensure smooth convergence at increased computational cost (such as SCF damping) for all fragment calculations,
which is computationally very expensive or instead monitor the SCF convergence for each specific case to automatically adapt these settings on the fly for a more efficient approach.

Finally, we point out that the fragmentation protocol presented in this section can cope with inorganic substructures in a molecular system. However, our rules for
which type of bonds are cleavable and how to saturate these must then be extended for molecular systems that comprise large inorganic substructures.
However, rather than extending the set of rules, it is more elegant to exploit embedding techniques that allow one to cut through covalent bonds
without a posterior valence saturation~\cite{manby12, fornace15, muhlbach18, lee19}.
Note that the modular structure of our approach easily allows for such extensions.

\subsection{Accuracy of the MM model}
\label{sec:accuracy_MM}

\begin{figure}[H]
\begin{center}
\vspace*{-0.5cm}
\includegraphics[width=0.65\textwidth,trim={0cm 0cm 0cm 0cm},clip]{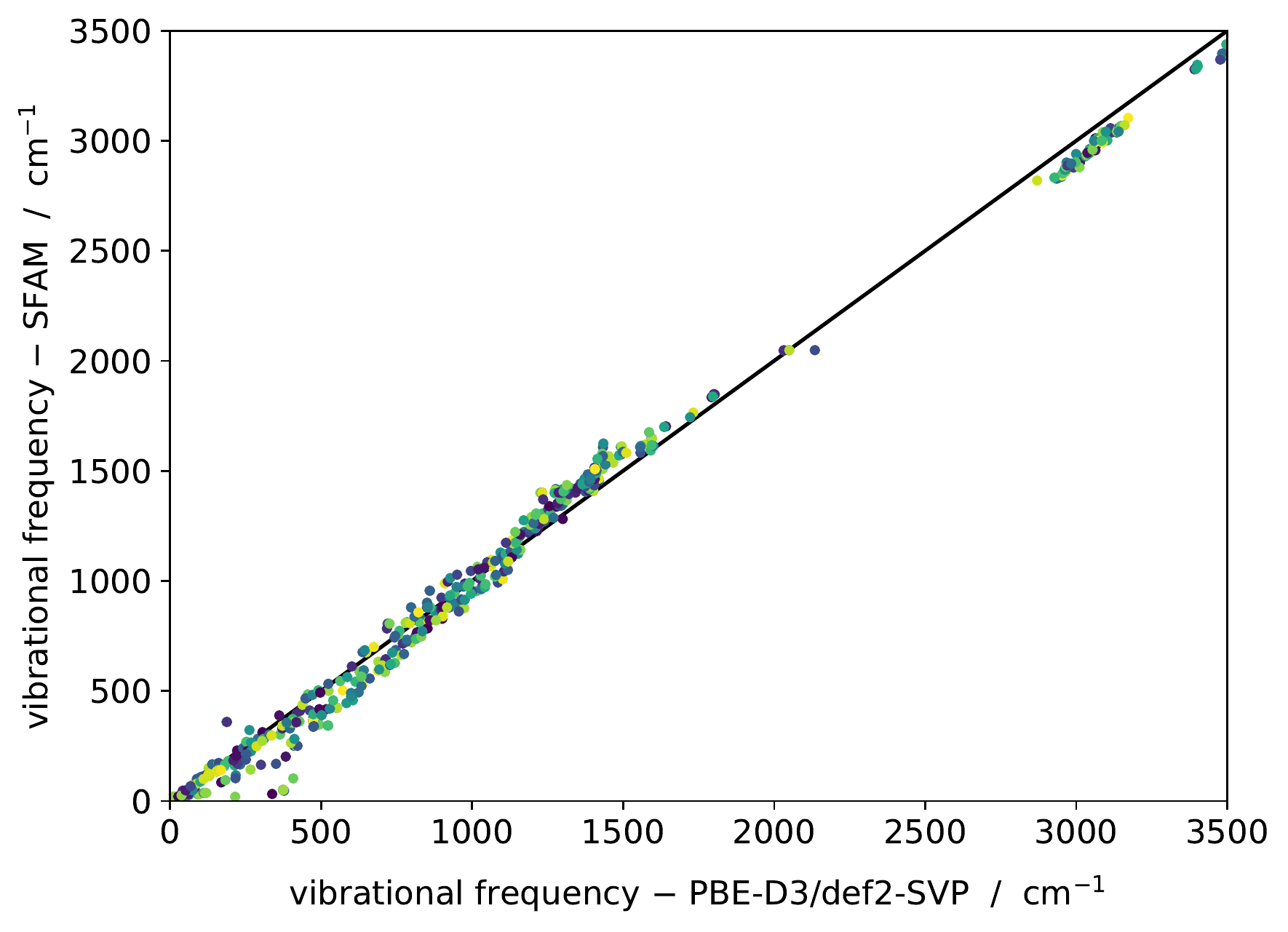}
%trim={<left> <lower> <right> <upper>
\end{center}
\vspace*{-0.8cm}
\caption{\label{fig:vibr_freq_comparison}\small Comparison of harmonic 
vibrational frequencies calculated for 11 small organic and inorganic molecules (see Supporting Information) obtained with PBE-D3/def2-SVP and SFAM, respectively.
Random colors are assigned to the dots for the sake of clarity.}
\end{figure}

We now assess the accuracy and the desired features of our automated MM parametri\-zation scheme.
First, the quality of the Hessian fit of the force constants is evaluated by comparing the SFAM calculated vibrational frequencies to their DFT reference.
As an example, we adopt the PBE-D3/def2-SVP~\cite{perdew96, grimme10, weigend05} electronic structure model as a reference.
CM5 charges were calculated with the \texttt{Gaussian09} software package~\cite{gaussian09}, all other electronic structure calculations were conducted with the \texttt{ORCA 4.1} program~\cite{neese12, neese18}.
A small set of 11 small and medium-sized organic and inorganic molecules (butane, hexane, aniline, alanine, methionine, tyrosine, chromium hexacarbonyl, tetramethylsilane, titanium tetrachloride,
melatonin, and anserine) was prepared and the detailed results are provided in the Supporting Information.
The mean absolute error (MAE) of the wavenumbers calculated by SFAM compared to their reference 
for the molecules in this set range between 12.7 and 116.6~cm$^{-1}$, which is in line with the
results obtained by Hirao and coworkers in their study~\cite{wang18} of the partial Hessian fitting approach.
The correlation between the SFAM frequencies and their reference values is presented in Fig.~\ref{fig:vibr_freq_comparison}.
The overall MAE for all 552~data points is 66.5~cm$^{-1}$ (however, the value obviously depends on the choice of the set of test molecules).
As expected, high frequency modes are described accurately, while larger deviations by up to 400~cm$^{-1}$ are observed in the low frequency spectrum,
because these data correspond to vibrational modes with little local character.
For these delocalized modes, this is to be expected because the approximation of describing molecular vibrations by uncoupled harmonic oscillators is unreliable.

\begin{figure}[H]
\begin{center}
\vspace*{-0.5cm}
\includegraphics[width=0.65\textwidth,trim={0cm 0cm 0cm 0cm},clip]{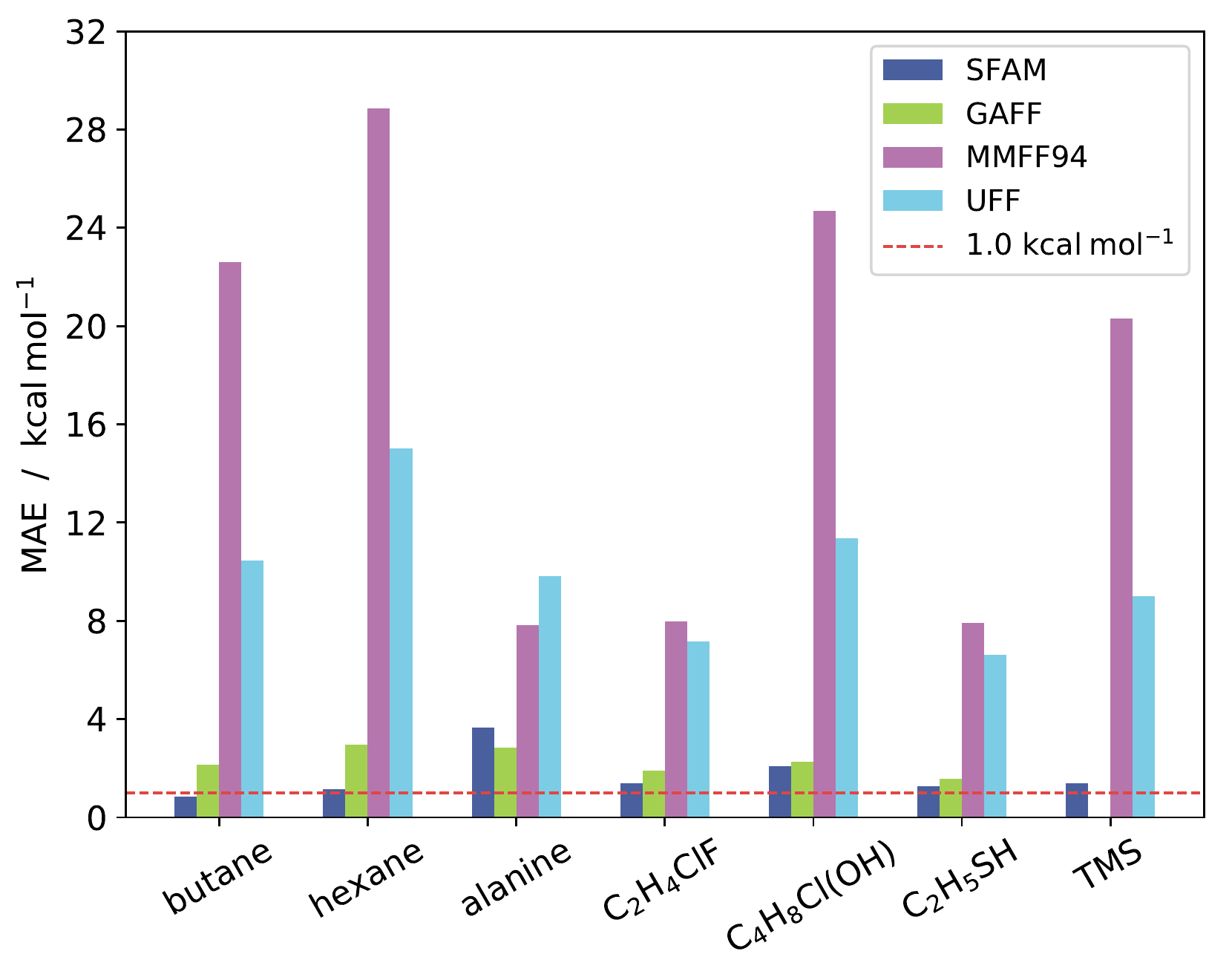}
%trim={<left> <lower> <right> <upper>
\end{center}
\vspace*{-0.8cm}
\caption{\label{fig:MD_energies_comparison}\small Comparison of MAE for SFAM relative energies according to
Eq.~(\ref{eq:MD_relative_energies}) 
for 10\,000 snapshots along a SFAM MD trajectory with three standard force field implementations from the Open Babel toolbox.
C$_2$H$_4$ClF and C$_4$H$_8$Cl(OH) refer to the isomers 4-chloro-1-butanol and 1-chloro-2-fluoroethane, respectively.
For tetramethylsilane (TMS), no result could be obtained for GAFF due to the absence of parameters.}
\end{figure}

Second, we investigate the performance of SFAM on molecular configurations from a broad range of the PES.
For this purpose, 10\,000 snapshots of a SFAM MD trajectory were taken
(details on the generation of the trajectory can be found in part~\ref{sec:ML} of this work).
For all of these structures, relative energies $\Delta E_i = E_i - E_{\text{eq}}$ were calculated with the reference method and SFAM,
as well as with three standard force fields (GAFF~\cite{wang04}, MMFF94~\cite{halgren96}, UFF~\cite{rappe92}) as implemented in the Open Babel cheminformatics toolbox~\cite{openbabel}.
We compare the MAE with respect to the relative reference energies $\Delta E_{\text{DFT},i}$,
\begin{equation}
 \text{MAE}_{\Delta E} = \frac{1}{n_s} \, \sum_{i\,=\,1}^{n_s} \, \lvert \, \Delta E_{\text{MM},i} - \Delta E_{\text{DFT},i} \, \rvert \quad , \label{eq:MD_relative_energies}
\end{equation}
for all four force fields and an arbitrary selection of seven molecules.
The results are depicted in Fig.~\ref{fig:MD_energies_comparison}.

In general, SFAM yields the smallest deviations from the reference DFT energies, which is remarkable as it was not
parametrized for arbitrary structures across a PES (however, GAFF performed better for the alanine molecule).
With the exception of the tetramethylsilane~(TMS) system for which no energies could be calculated due to the absence of parameters, one obtains reasonable energies for the GAFF model as well,
which is not surprising since GAFF was parametrized for small organic molecules.
By contrast, we observed much more severe deviations for MMFF94 and the Universal Force Field~(UFF).
We conclude that our automated parametrization scheme generates a model which describes the PES of the
test systems at least as well as other standard force fields.
Two main advantages are observed,
namely that a system-focused parametrization yields optimal force constants for the molecular system compensating for the lack of reference data for non-equilibrium configurations in the parameter optimization
and that, unlike in GAFF, naturally no restrictions on the type of molecular system emerge (cf. the TMS example).

However, we emphasize that this test is biased towards SFAM since the structures were sampled from a SFAM trajectory, i.e., structures closer to the equilibrium structure, on which SFAM has been parametrized,
were visited more frequently during the simulation. Removing this bias may be possible by reweighting the trajectories to the QM ensemble as discussed by York and coworkers~\cite{konig18_a}. Furthermore,
the MMFF94 and UFF approaches are not the most recent force fields in contrast to others that were listed in the Introduction, the former were chosen
because of their availability in the Open Babel toolbox. Hence, these two force fields do not necessarily represent the best specialized molecular mechanics models
currently available for our purpose.
However, this comparison is still valid for demonstrating our key point that
SFAM reaches comparable accuracy to standard force fields for typical molecular systems already without any machine learned corrections.
As explained above, additional investigations are necessary for a more detailed assessment of the SFAM base MM model compared to other well-established modern force fields,
at least for those molecule classes for which reliable standard force fields are available.

\begin{figure}[H]
\begin{center}
\vspace*{-0.5cm}
\includegraphics[width=0.65\textwidth,trim={0cm 0cm 0cm 0cm},clip]{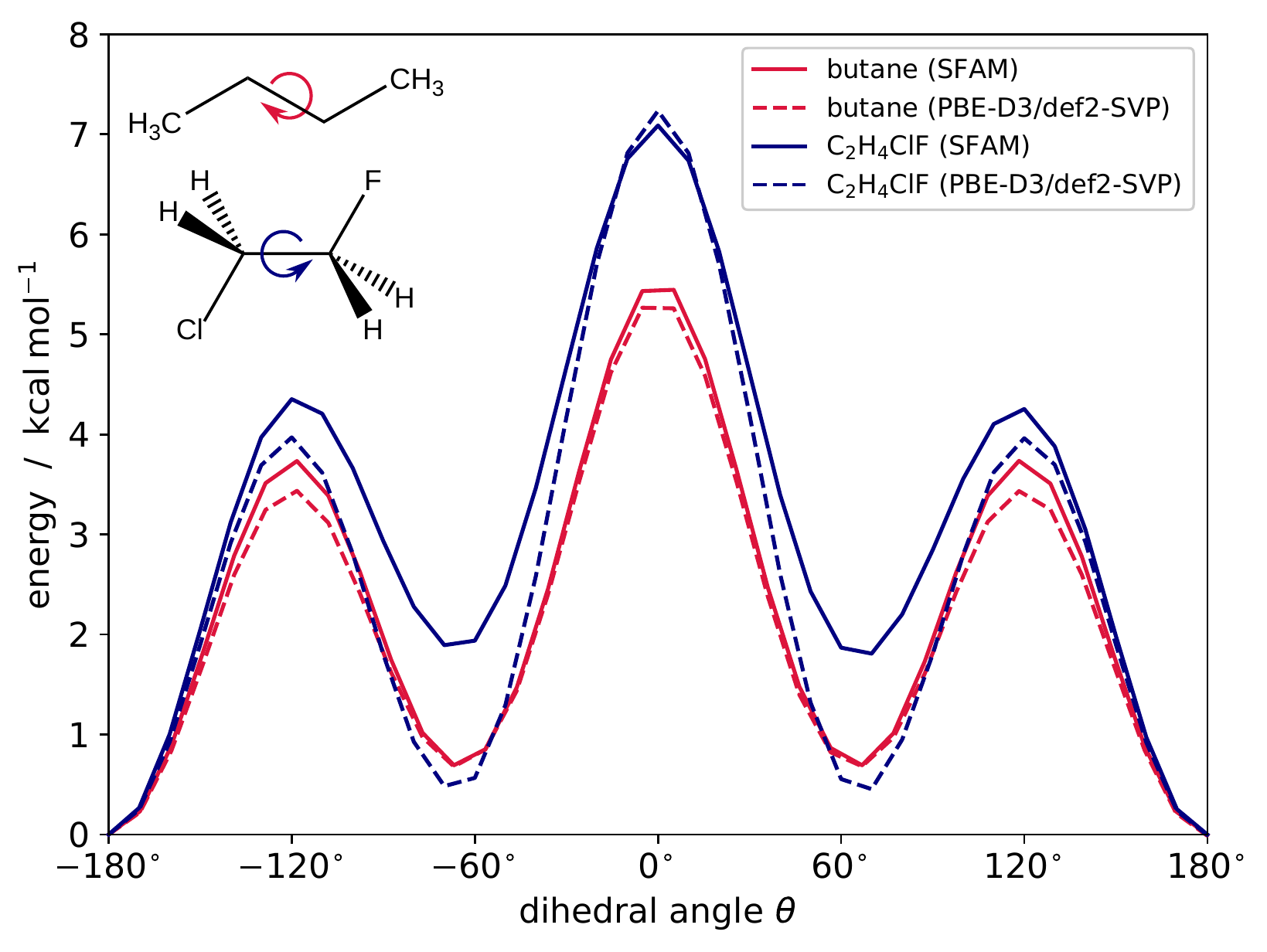}
%trim={<left> <lower> <right> <upper>
\end{center}
\vspace*{-0.8cm}
\caption{\label{fig:dihedral_angle_result_1}\small Complete $2\pi$ rotational scan around the central bond of butane (red) and 1-chloro-2-fluoroethane (blue) %C$_2$H$_4$ClF
calculated with SFAM and the DFT reference chosen for its parametrization. Both molecules were parametrized at their equilibrium 
structure,
which corresponds to $\theta=-180^{\circ}$. The energies of these structures were set to zero. The excellent agreement close to the equilibrium structures
and a decent description of the overall PES landscape for both cases is clearly visible.}
\end{figure}

\begin{figure}[H]
\begin{center}
\vspace*{-0.5cm}
 \includegraphics[width=0.65\textwidth]{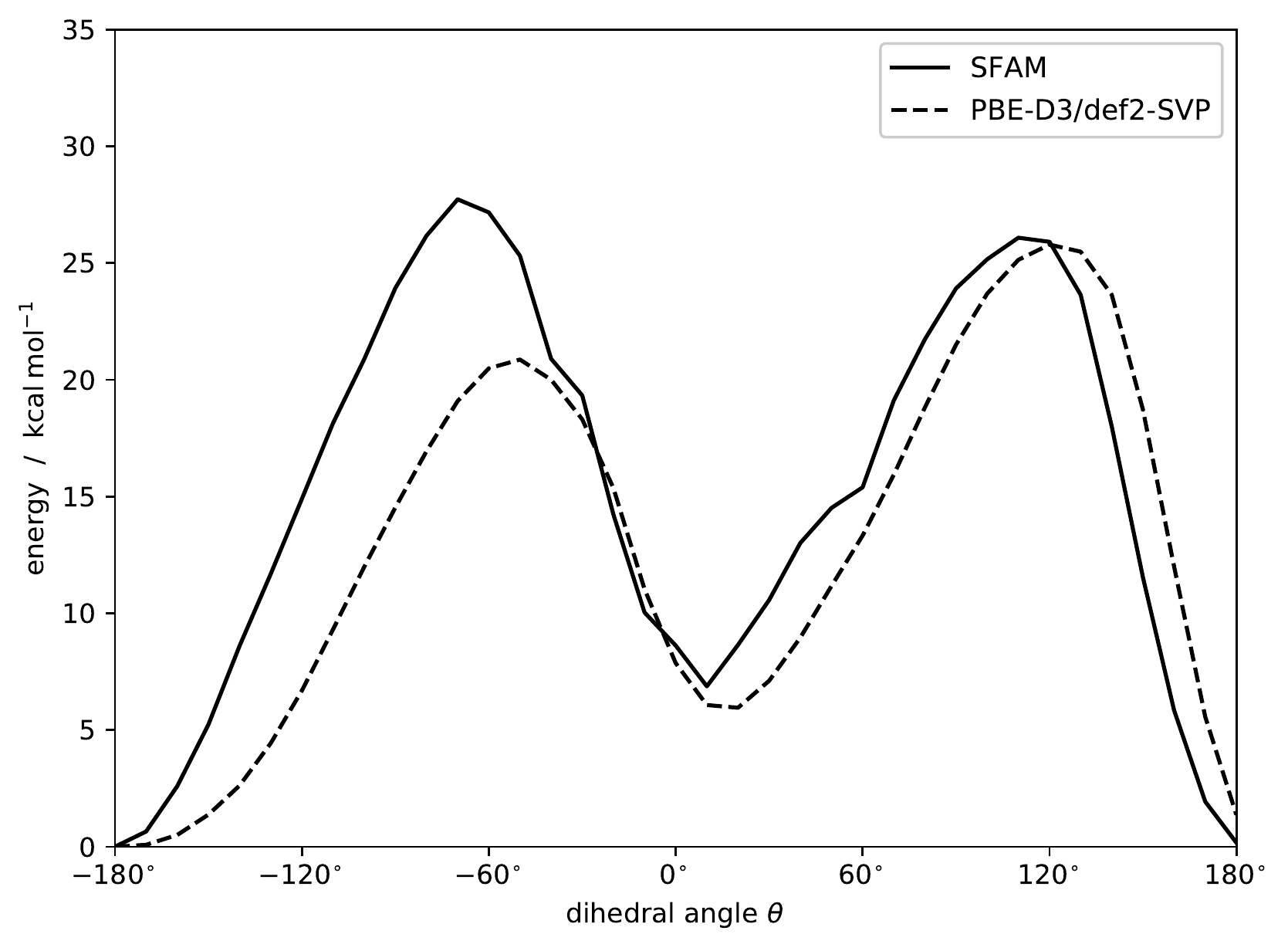}
\end{center}
\vspace*{-0.8cm}
 \caption{\label{fig:dihedral_angle_result_2}\small Complete 2$\pi$ rotational scan around the peptide bond in the dipeptide dialanine (alanyl-alanine) calculated with SFAM (solid line)
 and the DFT reference chosen for its parametrization (dashed line). The energies of the equilibrium structures ($\theta = -180^\circ$) were set to zero.}
\end{figure}

Finally, we investigate how dihedral potentials are described in SFAMs.
Fig.~\ref{fig:dihedral_angle_result_1} illustrates how the symmetric cosine potential fitted to partial Hessians
and the non-covalent interactions of the model combine to produce an asymmetric dihedral potential for two prototypical examples,
butane and 1-chloro-2-fluoroethane. In the case of butane, this leads to very good agreement with the reference method not only close to the equilibrium 
structure, but also for other values of the dihedral angle $\theta$.
In the case of the substituted ethane molecule, the potential landscape close to the equilibrium structure as well as to
the first barrier height are accurately described.
However, the relative energy of the local minima at $60^{\circ}$ and $-60^{\circ}$ with respect to the equilibrium 
structure deviates from the DFT reference by roughly 1.5~kcal\,mol$^{-1}$.
It is important to note, that the positions of the secondary minima are obtained correctly and that the essential features of the PES are recovered by SFAM.
To demonstrate that a similar behavior is observed in systems that are prototypical for biomolecular simulations,
we present the example of a dihedral surface scan of the dipeptide dialanine in Fig.~\ref{fig:dihedral_angle_result_2}.
This example also highlights the conclusion that SFAM is able to reproduce the positions of the energy minima as well as
the the essential features of the PES for peptide bond distortions.

To illlustrate the applied atom-type definitions explained in section~\ref{sec:atom_type_definitions},
Table~\ref{tab:atom_type_overview} gives an overview of the atom-type information of some of the systems studied in this section.

Having validated the key design features of SFAM, we now employ it as a base model for machine learned corrections
based on more reference data collected for non-equilibrium molecular configurations on the fly.
We emphasize, however, that already in its current form, SFAM can be applied in molecular dynamics simulations 
as it would be as reliable as a standard force field implementation, while not suffering
from any restrictions with respect to a chemical element that might occur in a molecule under consideration.

\vspace{0.5cm}
\begin{table}[H]
\renewcommand{\baselinestretch}{1.0}
\renewcommand{\arraystretch}{1.2}
\caption{\label{tab:atom_type_overview}\small Overview of the total number of atom types $N_\text{at,\,tot}$ and the resulting number $N_\text{p}$ of force constant parameters ($k_r$, $k_\alpha$, $k_\varphi$, $V_\theta$)
obtained for a selection of molecules with $N$ atoms. Additionally, we list the number of atom types for carbon atoms $N_\text{at,\,C}$ and for hydrogen atoms $N_\text{at,\,H}$, respectively.}
\begin{center}
\begin{tabular}{l l l l l l} 
\hline
\hline
molecular system \quad \quad \quad \quad & $N$ & $N_\text{at,\,tot}$ & $N_\text{at,\,C}$ & $N_\text{at,\,H}$ & $N_\text{p}$ \\
\hline 
butane & 14 & 4 & 2 & 2 & 12\\
hexane & 20 & 4 & 2 & 2 & 13\\
1-chloro-2-fluoroethane & 8 & 6 & 2 & 2 & 14\\
alanine & 13 & 10 & 3 & 4 & 27\\
dialanine & 23 & 13 & 4 & 5 & 13\\
methionine & 20 & 15 & 5 & 6 & 43\\
TiCl$_4$ & 5 & 2 & 0 & 0 & 2 \\
Cr(CO)$_6$ & 13 & 4 & 1 & 0 & 4\\
melatonin & 33 & 20 & 10 & 7 & 82\\
anserine & 33 & 24 & 9 & 9 & 81\\
plastocyanin & 1412 & 45 & 18 & 16 & 284\\
\hline
\end{tabular}
\renewcommand{\baselinestretch}{1.0}
\renewcommand{\arraystretch}{1.0}
\end{center}
\end{table}

%%%%%%%%%%%%%%%%%%%% MACHINE LEARNING %%%%%%%%%%%%%%%%%%%%
\section{On-the-fly model improvement with machine learning}
\label{sec:ML}

\subsection{Theory}
\label{sec:theory2}

Machine learning enables one to perform highly non-linear regression tasks to predict an output $y$ from a vector of inputs (or features) $\boldsymbol{x}$
based on training data points $\left\lbrace \tilde{\boldsymbol{x}}_i,\,\tilde{y}_i \right\rbrace$ with $i=1,2,3,\dots,m$.
In this work, we apply the efficient Kernel Ridge Regression (KRR) method~\cite{geron17}. It relies on a simple regularized linear regression model, which predicts a new data point $y$ according to
\begin{equation}
 y = \left( \mathbf{X}^{\text{T}} \mathbf{X} + \lambda \mathbf{I} \right)^{-1} \mathbf{X}^{\text{T}}\:\tilde{\boldsymbol{y}}\:\boldsymbol{x} \quad , \label{eq:linear_regression}
\end{equation}
where the matrix $\mathbf{X}$ holds all training data input $\tilde{\boldsymbol{x}}_i$, the vector $\tilde{\boldsymbol{y}}$ contains their target values,
and $\lambda$ scales the model regularization to avoid overfitting.
To apply this scheme to non-linear problems, the input space can be transformed into a higher-dimensional feature space.
A linear regression in this space corresponds to a non-linear regression in the original input space, but explicit 
calculations in such a space are typically unfeasible. 
However, Eq.~(\ref{eq:linear_regression}) can be kernelized, i.e., it can be written in a form that just relies on the 
calculation of dot products of inputs (\textit{kernel trick}).~\cite{scholkopf02}
A kernel is a function $K(\boldsymbol{x}_i,\boldsymbol{x}_j)$ that obtains
a dot product in a (possibly very high-dimensional) feature space. It does not require the knowledge of the transformation, just its existence.
The prediction of a new data point with KRR can be written as
\begin{equation}
 y = \sum_{i\,=\,1}^m \: \beta_i \: K(\tilde{\boldsymbol{x}}_i, \boldsymbol{x}) \quad ,
\end{equation}
with $K(\tilde{\boldsymbol{x}}_i, \boldsymbol{x})$ as the kernel and $\beta_i$ representing the new KRR regression weights, for which a closed expression can be derived from Eq.~(\ref{eq:linear_regression}).
Kernels exploited in this work comprise the linear kernel $K(\boldsymbol{x}_i, \boldsymbol{x}_j) = \boldsymbol{x}_i^{\text{T}} \: \boldsymbol{x}_j$,
the polynomial kernel $K(\boldsymbol{x}_i, \boldsymbol{x}_j) = \left( \gamma \boldsymbol{x}_i^{\text{T}} \: \boldsymbol{x}_j + c_0 \right)^d$ of degree $d$,
the Gaussian kernel $K(\boldsymbol{x}_i, \boldsymbol{x}_j) = \exp \left( -\frac{1}{2\sigma^2} \: \lvert \boldsymbol{x}_i - \boldsymbol{x}_j \rvert^2 \right)$,
and the Laplacian kernel $K(\boldsymbol{x}_i, \boldsymbol{x}_j) = \exp \left( -\frac{1}{\sigma} \: \left|\left| \boldsymbol{x}_i - \boldsymbol{x}_j \right|\right|_1 \right)$.

An estimate of the generalization error of an ML model, i.e., its performance on unseen data, can be obtained, for instance, from $k$-fold cross validation.~\cite{geron17}
In this algorithm, the data is divided into $k$ subsets. Within $k$ iterations, each subset is once set to be the test data, on which a model trained on the other $k-1$ subsets is evaluated.
The estimated performance, e.g., in terms of MAE, is obtained as the mean of all $k$ iterations, while the standard deviation can be taken as an uncertainty estimate.

\subsection{Methodology of learning energies and forces\\in molecular systems}
\label{sec:learning_energies_forces}

In this section, we describe a method to learn corrections to the energies and forces of molecular configurations calculated by our molecular mechanics model.
We emphasize that during this process none of the MM model parameters are re-evaluated individually, instead the resulting energies and forces of the model are assessed directly.
For comparison, the reference energies and forces are learned without a base model.
The energy of a certain molecular structure is a translationally and rotationally invariant scalar property.
We apply the well-known Coulomb matrix descriptor,\cite{rupp12}
\begin{equation} \label{eq:coulomb_matrix}
 \mathbf{M}_{AB} =  \begin{cases}
                 0.5Z_A^{2.4} & \text{for} \: A = B \\
                 \frac{\displaystyle Z_AZ_B}{\displaystyle \lvert \boldsymbol{r}_A - \boldsymbol{r}_B \rvert} & \text{for} \: A \neq B \quad ,
                \end{cases}
\end{equation}
with the aforementioned properties.
$Z_A$ represents the nuclear charge and $\boldsymbol{r}_A$ the Cartesian coordinates of atom $A$.
The ordering of the atom indices is kept consistent for all molecular configurations.
Multiple ways exist to extract a set of features for the ML algorithm from this representation.~\cite{rupp15review,hansen13} 
We construct our feature vector as the vectorized upper triangle of $\mathbf{M}$.

By contrast, forces are vector-valued quantities, whose Cartesian representation depends on the absolute orientation of the molecular system.
Therefore, the straightforward approach of Eq.~(\ref{eq:coulomb_matrix}) will not yield a working model to predict forces.
Two conditions are required to set up a well performing ML model: i) the molecular representation needs to contain directional information
and ii) the atomic forces shall be learned in an internal coordinate system, which is not dependent on the molecular orientation in the global reference frame.
Several approaches addressing this challenge have been discussed in the literature.~\cite{chmiela17, botu15, botu15ijqc, li15, suzuki17, rupp15}
Inspired by the work of von Lilienfeld and coworkers~\cite{rupp15}, we apply the following procedure.
To define an atom-centered internal coordinate system for an atom $A$, a rectangular matrix $\mathbf{X}_A$ is constructed to represent its chemical environment,
\begin{equation} \label{eq:pca_input_matrix}
  \mathbf{X}_A =
 \begin{pmatrix}
  \displaystyle\frac{Z_1 \left( x_1 - x_A \right)}{\lvert\boldsymbol{r}_1 - \boldsymbol{r}_A \rvert^3} &  \displaystyle\frac{Z_1 \left( y_1 - y_A \right)}{\lvert\boldsymbol{r}_1 - \boldsymbol{r}_A \rvert^3} & \displaystyle\frac{Z_1 \left( z_1 - z_A \right)}{\lvert\boldsymbol{r}_1 - \boldsymbol{r}_A \rvert^3} \\
  \vdots & \vdots & \vdots \\
  \displaystyle\frac{Z_M \left( x_M - x_A \right)}{\lvert\boldsymbol{r}_M - \boldsymbol{r}_A \rvert^3} &  \displaystyle\frac{Z_M \left( y_M - y_A \right)}{\lvert\boldsymbol{r}_M - \boldsymbol{r}_A \rvert^3} & \displaystyle\frac{Z_M \left( z_M - z_A \right)}{\lvert\boldsymbol{r}_M - \boldsymbol{r}_A \rvert^3}
 \end{pmatrix}
 \quad ,
\end{equation}
containing row-wise information about its $M$ neighbor atoms within a distance threshold $\tilde{r}$, favoring heavy atoms and decreasing the influence of distant atoms.
A Principal Component Analysis (PCA) of $\mathbf{X}_A$ yields its three principal components $\boldsymbol{c}$, which represent the three basis vectors of our atom-centered internal coordinate system for atom $A$.
The transformation matrix $\mathbf{P}$, with its columns holding the principal component vectors $\boldsymbol{c}$, can be applied to transform the Cartesian representation of the force vector $\boldsymbol{f}_A$
to its internal representation $\boldsymbol{f}_A^{\text{int}}$,
\renewcommand\arraystretch{0.8}
\begin{equation} \label{eq:forces_transformation}
 \boldsymbol{f}_A^{\text{int}} = \mathbf{P}^{-1} \boldsymbol{f}_A =
 \begin{pmatrix}
    \vert & \vert & \vert \\
    \boldsymbol{c}_1   & \boldsymbol{c}_2 & \boldsymbol{c}_3   \\
    \vert & \vert & \vert 
\end{pmatrix}
^{-1} \boldsymbol{f}_A
\quad ,
\end{equation}
and vice versa, $\boldsymbol{f}_A = \mathbf{P}\boldsymbol{f}_A^{\text{int}}$.
We point out that the components are only unique up to a sign, hence we adopt the convention introduced by von Lilienfeld and coworkers~\cite{rupp15},
where the sign of the PCA axes are chosen in such a way that the center of nuclear charge of the $M$ neighbors has positive coordinates only.
The feature matrix $\mathbf{M}_A$ for atom $A$ is constructed analogously to Eq.~(\ref{eq:pca_input_matrix}),
\renewcommand\arraystretch{1.5}
\begin{equation} 
  \mathbf{M}_A =
 \begin{pmatrix}
  \displaystyle\frac{Z_1 x_1^{\text{int}}}{\lvert\boldsymbol{r}_1^{\text{int}}\rvert^3} &  \displaystyle\frac{Z_1 y_1^{\text{int}}}{\lvert\boldsymbol{r}_1^{\text{int}}\rvert^3} & \displaystyle\frac{Z_1 z_1^{\text{int}}}{\lvert\boldsymbol{r}_1^{\text{int}}\rvert^3} \\
  \vdots & \vdots & \vdots \\
  \displaystyle\frac{Z_M x_1^{\text{int}}}{\lvert\boldsymbol{r}_M^{\text{int}}\rvert^3} &  \displaystyle\frac{Z_M y_1^{\text{int}}}{\lvert\boldsymbol{r}_M^{\text{int}}\rvert^3} & \displaystyle\frac{Z_M z_1^{\text{int}}}{\lvert\boldsymbol{r}_M^{\text{int}}\rvert^3}
 \end{pmatrix}
 \quad ,
\end{equation}
and the rows are ordered by the distance of their corresponding atoms to the central atom $A$.
Here, $\boldsymbol{r}_i^{\text{int}}$ is the representation of the Cartesian coordinates in the internal coordinate system centered around atom $A$,
\begin{equation}
\boldsymbol{r}_i^{\text{int}} = \mathbf{P}^{-1} \left( \boldsymbol{r}_i - \boldsymbol{r}_A \right)
 \quad ,
\end{equation}
hence,
the norm of this vector equals the distance between atom $i$ and atom $A$.
To ensure a fixed size of $\mathbf{M}_A$ for all molecular configurations, the decision on which atoms appear in $\mathbf{M}_A$,
i.e., the ones closer to $A$ than the threshold $\tilde{r}$, is based on one fixed molecular configuration, for instance, the equilibrium molecular structure.

The aforementioned method is applied to each atom individually.
All atomic forces are transformed into their internal representation by Eq.~(\ref{eq:forces_transformation}) for training and prediction.
To make the forces applicable, e.g., in a molecular dynamics simulation, the predicted vectors for each atom can easily be transformed back into their Cartesian representation.

\subsection{Results}
\label{sec:results2}

\begin{figure}[H]
\begin{center}
\vspace*{-0.5cm}
\includegraphics[width=0.65\textwidth,trim={0cm 0cm 0cm 0cm},clip]{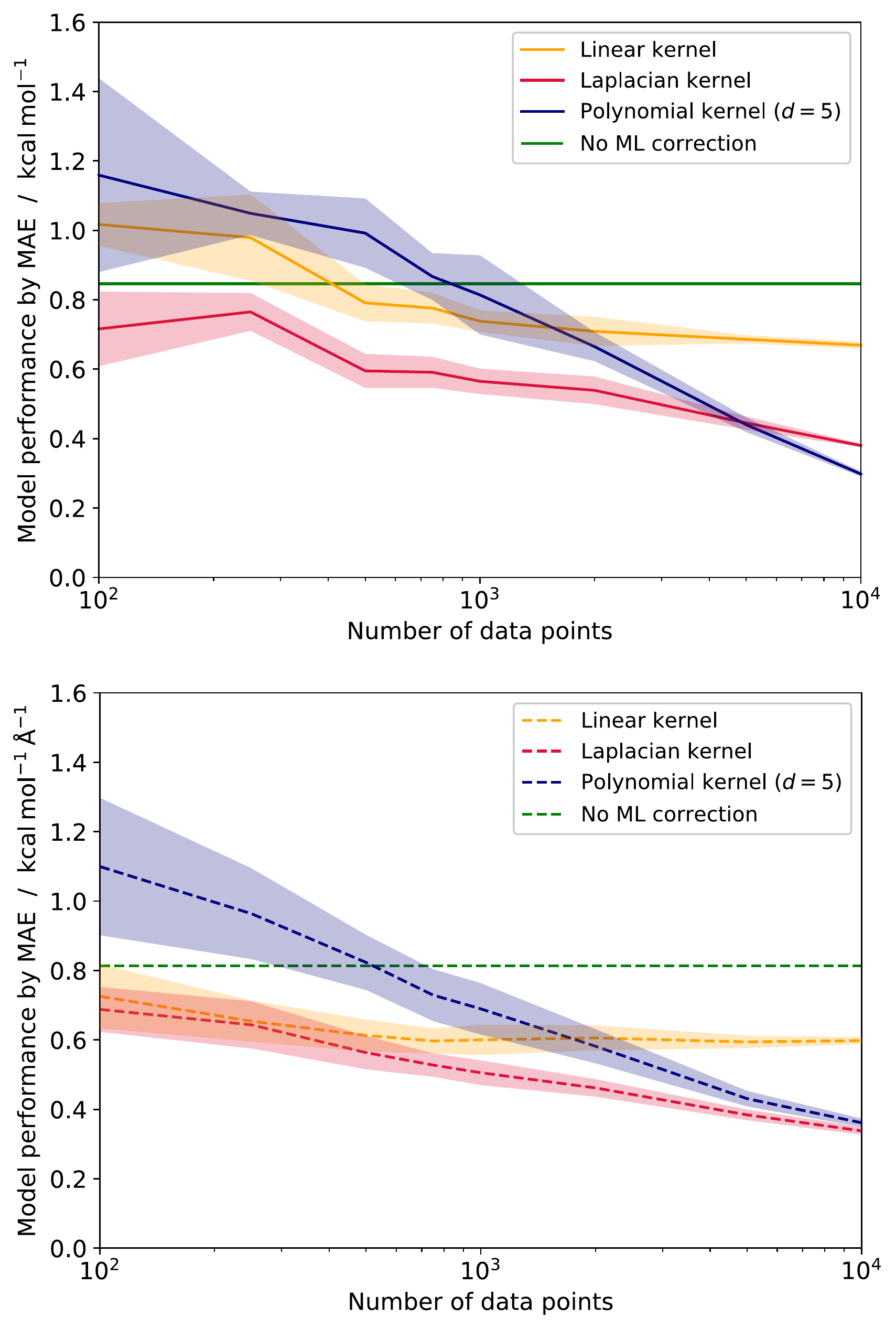}
%trim={<left> <lower> <right> <upper>
\end{center}
\vspace*{-0.5cm}
\caption{\label{fig:butane_hybrid_ML}\small Evaluation of the MM/ML hybrid model for electronic energies (top) and forces (bottom)
of butane configurations by 5-fold cross validation for differently sized data sets.
Each data point corresponds to a molecular structure for which a reference calculation is available.
The shaded areas represent a confidence interval of one standard deviation.}
\end{figure}

\begin{figure}[H]
\begin{center}
\vspace*{-0.5cm}
\includegraphics[width=0.65\textwidth,trim={0cm 0cm 0cm 0cm},clip]{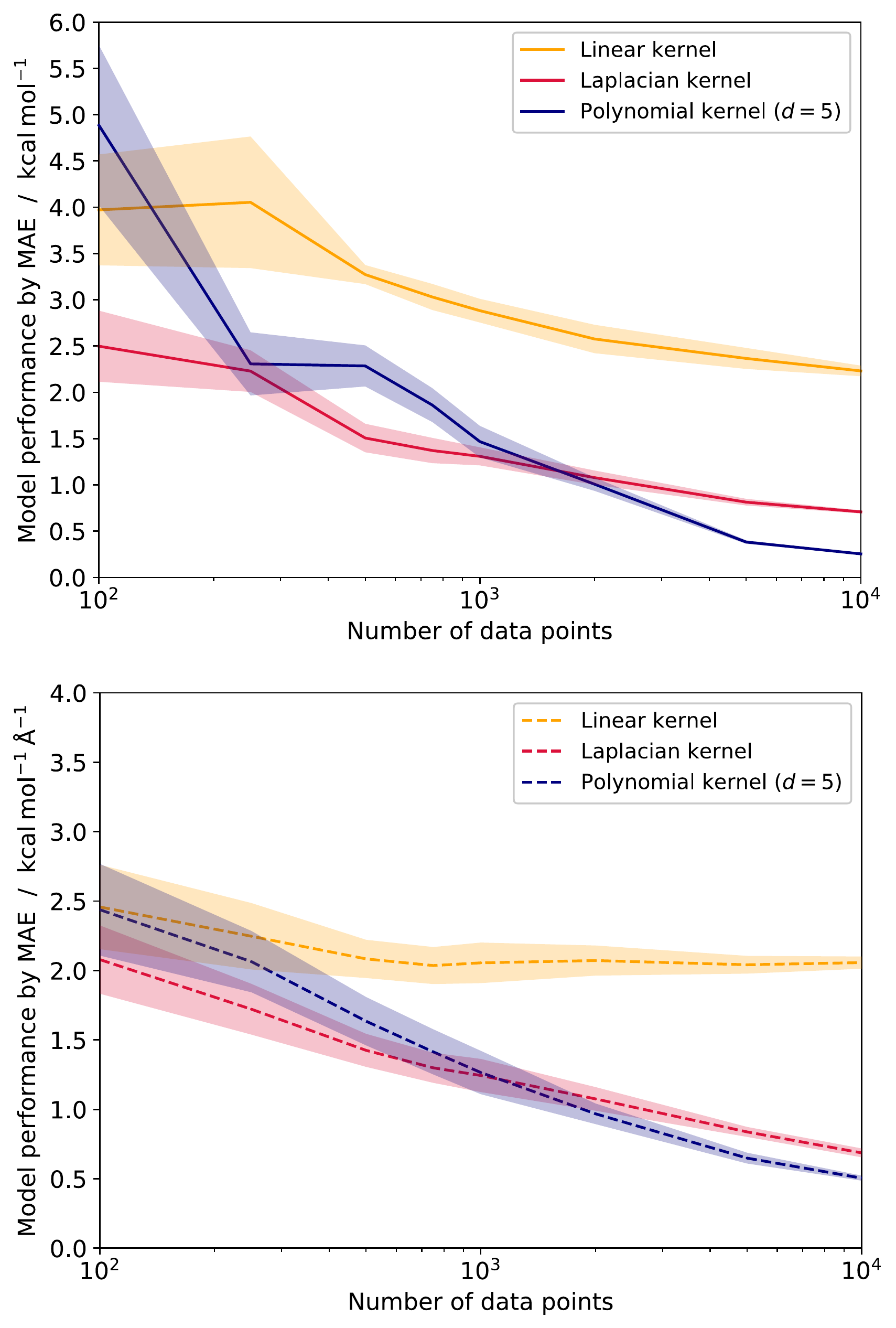}
%trim={<left> <lower> <right> <upper>
\end{center}
\vspace*{-0.5cm}
\caption{\label{fig:butane_only_ML}\small Evaluation of an ML-only model for electronic energies and forces of butane configurations by 5-fold cross validation for differently sized data sets.
Each data point corresponds to a molecular structure for which a reference calculation is available.
The shaded areas represent a confidence interval of one standard deviation.}
\end{figure}

We demonstrate the capabilities of our hybrid MM/ML method for the butane molecule. More examples are presented in the Supporting Information.
First, an MM model is parametrized for butane automatically from a PBE-D3/def2-SVP optimized structure and its PBE-D3/def2-SVP Hessian matrix.
All other model parameters were determined as described in part~\ref{sec:automated_mm_parametrization}.
With this model, an MD trajectory of the molecule in a vacuum was generated employing the leap-frog integration scheme at a constant temperature of 300~K with a timestep of 1~fs.
After reaching thermal equilibrium, 10\,000 structures were taken from the trajectory at time intervals of 2.5~ps.
For each of the structures, reference energies and forces were calculated with the same quantum chemical reference method employed for the initial parametrization.
Differently sized subsets were extracted to evaluate the performance of the KRR with varying training set size.
All kernels mentioned in section~\ref{sec:theory2} were investigated, 
but results for only three of them are displayed in Figs.~\ref{fig:butane_hybrid_ML} and \ref{fig:butane_only_ML} for clarity.
We chose the linear kernel for its simplicity, the polynomial kernel of fifth degree for the flexibility of the resulting model,
and the Laplacian kernel was added due to its previously reported outstanding results in similar prediction tasks~\cite{hansen13}.

For the sake of comparison, we trained and evaluated one ML model for the differences of our base model to the reference values ($\Delta$-ML) and one ML model, which learns the reference energies and forces directly.
We exploit the given reference data to estimate the generalization error of the ML and MM/ML hybrid methods, i.e., the expected error on unseen molecular configurations,
which measures the model's performance.
As described in section~\ref{sec:theory2},
we apply $k$-fold cross validation (here, $k=5$) to obtain this estimate and the uncertainty connected to it.
Note that for 5-fold cross validation a data set size of $N$ corresponds to $0.8N$ training points.
The hyperparameters of the ML model were also determined by applying this technique.
In all figures, solid lines represent the ML predictions for electronic energies and dashed lines are chosen for atomic force predictions.

Fig.~\ref{fig:butane_hybrid_ML} demonstrates the reliability
of the MM/ML hybrid model for three given kernels in comparison to the pure MM base model for the energy and forces.
As expected, the model fidelity increases gradually with the number of available training data points, reaching mean absolute errors of below 0.4~kcal\,mol$^{-1}$ for very large training data sets.
Furthermore, a large amount of data enables a model evaluation with small uncertainties, which are estimated by the standard deviation obtained from a 5-fold cross validation.
Note, that the small choice of $k$ may result in an irregular behavior of the uncertainty estimates as seen in 
Figs.~\ref{fig:butane_hybrid_ML} and \ref{fig:butane_only_ML},
which can be improved by increasing $k$ to obtain more accurate estimates of the generalization error.
Applying the Laplacian kernel yields satisfactory corrections even for a small amount of reference data.
However, when accumulating a lot of reference data for a molecular system, it is advisable to switch to a model,
which revealed a steeper learning curve in our study, for instance, a polynomial kernel of third or fifth degree.
In general, the flexibility of the ML model should be tailored toward the amount of reference data available
by choosing an appropriate kernel or by modification of the model's hyperparameters.

Fig.~\ref{fig:butane_only_ML} shows the performance of an ML model learning the energy and forces directly with no MM baseline.
We identify a very similar learning performance for this case.
However, more than 1000 reference calculations are needed to reach chemical accuracy of approximately 1~kcal\,mol$^{-1}$ or 1~kcal\,mol$^{-1}$\,\AA$^{-1}$,
while the hybrid model already starts below this threshold or, in other examples (see Supporting Information), reaches it with a much smaller
amount of training data, whereas the exact amount strongly depends on the size and conformational flexibility of the molecular system.

Therefore, we observe several advantages of the hybrid approach. First,
having a quickly obtainable base model allows for much higher model accuracy for
limited amounts of reference data from sampled molecular configurations.
We parametrize our base model solely on single-point data (optimized structures and Hessians) that must be calculated anyways to verify that a molecular structure is a local energy minimum,
and hence, a potentially relevant chemical structure. Such small amount of data would not be sufficient for a pure machine learning model.
Second, the base model is essential to sample configurations initially in a consistent
manner. Moreover, the MM model can even be applied for computational studies
without any machine learned corrections.
Although small amounts of additional reference data may not increase the model accuracy substantially in every case,
they are already beneficial to provide an uncertainty estimate of the base model, which
may be refined by including even more data in a rolling fashion as needed.

Finally, we point out that even in cases where large amounts of reference data are available, starting out from a physically motivated MM model is beneficial, because (i) it allows for direct physical insight
into the system's properties compared to a pure machine learning approach, (ii) it provides an analytic-model representation that can already reliably cover a large fraction of the relevant configuration space
and (iii) its parametrization is easily scalable toward huge molecular frameworks (as demostrated in section~\ref{sec:subsystem_parametrization}).\\
We may also refer to recent work by Cs\'anyi and coworkers~\cite{oord19} who have also demonstrated that the explicit decomposition of the potential energy in terms of atomic many-body expansions
(which can be considered a generalization of the standard molecular mechanics model employed in our work)
reduces the curse of dimensionality for the parametrization of interatomic potentials compared to high-dimensional machine learning models and leads to better transferability of the model.

Note that an alternative to classical MD for generating structures from the relevant configuration space is offered by Monte Carlo simulations~\cite{binder95, frenkel01}.
This option may be expected to yield a more diverse set of structures that would require additional precautions in the MD simulations (such as restraints or high temperature).

\section{Conclusions}
\label{sec:conclusions}

\begin{figure}[H]
\begin{center}
\includegraphics[width=0.85\textwidth,trim={0cm 0cm 0cm 0cm},clip]{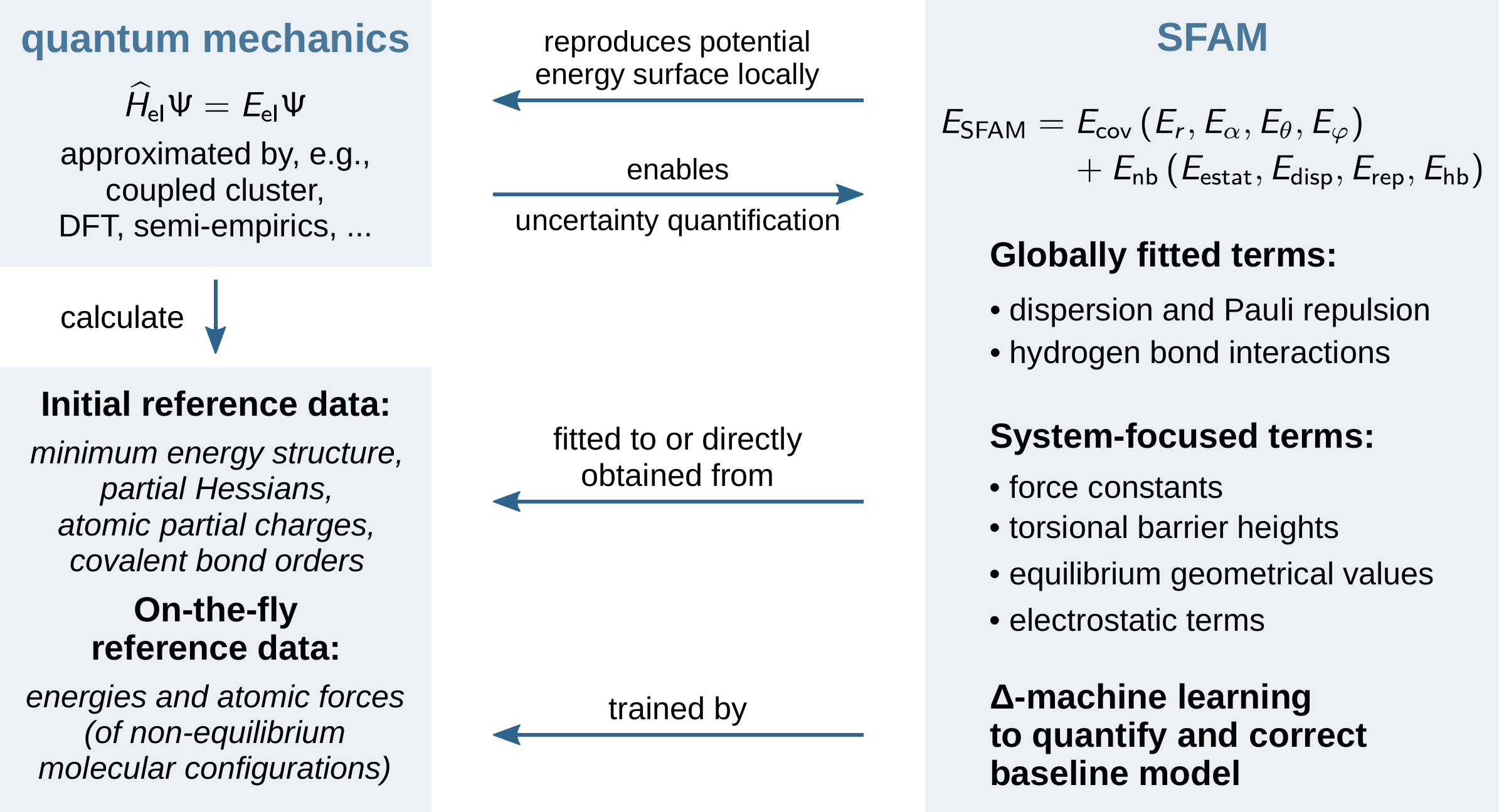}
%trim={<left> <lower> <right> <upper>
\end{center}
\vspace*{-0.25cm}
\caption{\label{fig:conclusion}\small Conceptual view of the SFAM approach.}
\end{figure}

We presented a strategy to obtain efficient self-parametrizing system-focused atomistic models~(SFAMs) from a combination of automatically generated physical models
and on-the-fly machine learned corrections including uncertainty quantification. Our SFAM approach is summarized in Fig.~\ref{fig:conclusion}.
The benefits of this approach can be summarized as follows. On the one hand, a physically motivated atomistic model
can be parametrized by small amounts of reference data without human interference.
The first-principles reference allows us to either apply a hierarchy of electronic structure methods of increasing accuracy or to exploit a given approach (such as DFT) with errors assigned.
On the other hand, data-driven models capable of learning highly non-linear relationships can compensate deficiencies of the approximate physical model
and additionally allow us to quantify the model's reliability.

We analyzed the SFAM ansatz for small organic and inorganic molecules and showed that the
choice of a partial-Hessian based parameter optimization does not limit the automated generation of MM models for large systems.
Furthermore, we outlined a molecular fragmentation scheme for this purpose.
As a result, we obtained an autonomous model construction scheme that can be applied to arbitrary nanoscale structures or even employed to set up such atomistic structures.
However, more testing will be required to understand the reliability and the robustness of SFAM in greater detail.
Our implementation will be made available in the context of the open-source \texttt{SCINE} project~\cite{scine}.

To work with a system-focused
model, such as a reduced-dimensional MM model obtained from full-dimensional QM description of the elementary particles of a reference 
molecular structure, in a quantitative fashion necessitates uncertainty quantification that measures the reliability of the model for
the specific case under consideration.
Following our previous work~\cite{simm18_gp, proppe19}, we exploit confidence intervals of machine learning schemes for this task.

Large systems are not only plagued by the unfavorable
scaling of electronic structure methods with system size, they will also suffer from
the exponentially increasing number of molecular conformations.
Therefore, applying machine learning models to a nanoscale system requires these to be based on small molecular fragments 
(embedded in the full system) to generate
reference data on the fly and in a reasonably short time frame. 

We note that the parametrized models produced with our approach may be collected in a universal centralized database~(in analogy to the PDB~\cite{bernstein77}
and similar to the QCArchive database~\cite{qcarchive} of the MolSSI project~\cite{wilkins18}) for future application of meta-ML methods to extract transferable model information.

In future work, we will develop our automated framework to allow for straightforwardly setting up QM/MM/ML calculations 
with uncertainty quantification for application in, for instance,
metalloenzyme or heterogeneous catalysis, where the accurate description of chemical reactions at local reactive sites within nanoscale systems 
benefits from an explicit quantum mechanical description.

%%%%%%%%%%%%%%%%%%%% ACKNOWLEDGEMENTS %%%%%%%%%%%%%%%%%%%%
\section*{Acknowledgments}
\label{sec:acknowledgments}
The authors thank the Schweizerischer Nationalfonds for generous support 
(Projects 200021\_182400 to M.R. and 200021\_172950-1 (C.B.) to PD Dr. Thomas Hofstetter).
C.B. gratefully acknowledges support by a Kekul\'{e} PhD fellowship of the Fonds der Chemischen Industrie.
This work was presented at the 9$^{\text{th}}$ Molecular Quantum Mechanics Conference (MQM 2019) in Heidelberg, Germany in July 2019.

\section*{Associated Content}
\label{sec:SI}
Supporting Information is provided containing detailed information about the globally fitted MM parameters,
the test set employed to evaluate the accuracy of SFAM on vibrational frequencies and the fragment sizes obtained
by the automated fragmentation protocol. Furthermore, we provide results of the MD simulation of plastocyanin
as well as two additional examples for on-the-fly model improvement with machine learning.
This information is available free of charge via the Internet at \texttt{http://pubs.acs.org}.

%%%%%%%%%%%%%%%%%%%% REFERENCES %%%%%%%%%%%%%%%%%%%%
%% \bibliographystyle{achemso2}
%\bibliography{references}

\begin{mcitethebibliography}{163}
\bibitem[Duster \latin{et~al.}(2017)Duster, Wang, Garza, Miller, and
  Lin]{duster17}
Duster,~A.~W.; Wang,~C.-H.; Garza,~C.~M.; Miller,~D.~E.; Lin,~H. {Adaptive
  quantum/molecular mechanics: What have we learned, where are we, and where do
  we go from here?} \emph{WIREs Comput. Mol. Sci.} \textbf{2017}, \emph{7},
  e1310\relax
\mciteBstWouldAddEndPuncttrue
\mciteSetBstMidEndSepPunct{\mcitedefaultmidpunct}
{\mcitedefaultendpunct}{\mcitedefaultseppunct}\relax
\EndOfBibitem
\end{mcitethebibliography}
\providecommand{\latin}[1]{#1}
\makeatletter
\providecommand{\doi}
  {\begingroup\let\do\@makeother\dospecials
  \catcode`\{=1 \catcode`\}=2 \doi@aux}
\providecommand{\doi@aux}[1]{\endgroup\texttt{#1}}
\makeatother
\providecommand*\mcitethebibliography{\thebibliography}
\csname @ifundefined\endcsname{endmcitethebibliography}
  {\let\endmcitethebibliography\endthebibliography}{}

% \newpage
%
% \begin{center}
% \textbf{Table-of-Contents Figure}
% 
%\includegraphics[width=8cm]{toc.pdf}
% \end{center}

\end{document}